\documentclass[11pt,a4paper]{article}

\usepackage{jheppub}
\usepackage[svgnames]{xcolor}
\usepackage{tikz}
\tikzset{>=latex}
\tikzset{baseline=(A.base)}
\tikzset{arbitrary/.style={}}
\tikzset{1gon/.style={draw}}
\usetikzlibrary{shapes.geometric}
\tikzset{SU/.style={draw,circle}}
\tikzset{Sp/.style={draw,semicircle,inner sep=2pt,shape border rotate=180}}
\tikzset{SO/.style={draw,semicircle,inner sep=2pt}}
\tikzset{Sq/.style={draw,rectangle,inner sep=5pt}}
\tikzset{2/.style={draw,regular polygon, regular polygon sides=3,minimum size=0.5cm,inner sep=0,shape border rotate=180}}
\tikzset{S/.style={draw,regular polygon, regular polygon sides=3,minimum size=0.5cm,inner sep=0}}

\def\SO{\mathrm{SO}}
\def\SU{\mathrm{SU}}
\def\USp{\mathrm{USp}}

\def\cN{\mathcal{N}}

\def\fund{\mathbf{fund}}
\def\vect{\mathbf{vect}}
\def\adj{\mathbf{adj}}
\def\asym{\mathbf{asym}}
\def\asymT{\mathbf{asym3}}
\def\sym{\mathbf{sym}}
\def\spinor{\mathbf{S}}
\def\conj{\mathbf{C}}
\def\half{\mathbf{\tfrac12}}

\title{Classification of 6d $\cN{=}(1,0)$ gauge theories}

\preprint{}

\author[1]{Lakshya Bhardwaj}
\affiliation[1]{Perimeter Institute for Theoretical Physics, Waterloo, Ontario, Canada N2L 2Y5}
\emailAdd{lbhardwaj@perimeterinstitute.ca}

\abstract{
We delineate a procedure to classify 6d $\cN=(1,0)$ gauge theories composed, in part, of a semi-simple gauge group and hypermultiplets. We classify these theories by requiring that they satisfy some consistency conditions. The primary consistency condition is that the gauge anomaly can be cancelled by adding tensor multiplets which couple to the gauge fields by acting as sources of instanton strings. Based on the number of tensor multiplets required to cancel the anomaly, we conjecture that the UV completion of these consistent gauge theories (if it exists) should be either a 6d $\cN=(1,0)$ SCFT or a 6d $\cN=(1,0)$ little string theory.
}


\begin{document}
\setcounter{tocdepth}{2}
\maketitle
\section{Introduction}
Weakly coupled gauge theories with conventional matter don't seem to be UV complete in 6d. Indeed, consider such a theory defined at some energy scale. Since the mass dimension of any gauge coupling $g$ in 6d is -1, the strength of the interactions becomes stronger and stronger in the UV and the theory lacks a proper definition.

However, in the mid-1990s, many examples of 6d $\cN=(1,0)$ SCFTs were found using brane constructions in string theory \cite{Witten:1995zh,Strominger:1995ac,Seiberg:1996vs,Hanany:1997gh}. Some of these theories have a tensor branch of vacua on which the effective theory in the IR becomes a conventional 6d $\cN=(1,0)$ gauge theory coupled to tensor multiplets. The number of tensor multiplets coincides with the number of gauge groups and the gauge couplings are controlled by the vevs of the scalars in the tensor multiplet \cite{Seiberg:1996qx}. The IR effective gauge theory has BPS instantons which are strings in 6d. The tension of these strings is also controlled by tensor moduli. At the origin of tensor branch, all of these strings become tensionless.

Moreover, similar constructions led to the discovery of new types of theories in 6d which enjoyed T-duality but did not have gravitons \cite{Seiberg:1997zk,Berkooz:1997cq}. These theories were termed as little string theories. Some of these theories also lead to a gauge theory in the IR but with one less tensor multiplet than the number of gauge groups. So, one of the gauge couplings remains a dimensionful parameter of the theory independent of the tensor moduli. Hence, these theories have a mass scale and are not conformal. This mass scale can also be viewed as the tension of a BPS instanton string. This string gives the necessary winding modes upon compactification that are required for T-duality.

A couple of remarks are in order to clear some confusions that may arise in above mentioned statements. First, in this paper, we are only considering 6d theories in which all the strings of the theory have an interpretation in terms of instanton strings of the low energy gauge theory on a generic point on tensor branch. In general, we should check if the number of tensor multiplets is equal to or one less than the number of strings in the theory to characterize if we have an SCFT or a little string theory, respectively. But in the class of theories we are considering, we can identify the number of strings with the number of simple gauge groups. Second, when we say that the number of tensor multiplets is equal to or one less than the number of gauge groups, we have in our mind \emph{irreducible} or \emph{connected} theories. Of course, one can consider many irreducible theories at the same time such that they are all decoupled from each other. If we consider a theory  which has $p$ irreducible little string theory sectors, then it would have number of tensor multiplets $p$ less than the number of gauge groups. From now on, we will only talk about connected theories.

Now, one can ask what are the allowed 6d $\cN=(1,0)$ gauge theories with conventional matter which can possibly arise as an IR effective field theory of a UV complete theory in the form of a 6d $\cN=(1,0)$ SCFT or little string theory. These gauge theories have to satisfy some consistency conditions which we will describe below. The purpose of this note is to outline the classification of all gauge theories satisfying these consistency conditions. We don't know if all of these gauge theories actually have a UV completion in the form of a 6d SCFT or little string theory. We find two possibilities:
\begin{itemize}
	\item The number of tensor multiplets required to cancel the gauge anomaly (via Green Schwarz mechanism) is equal to the number of gauge groups. There seems to be no mass parameter in these theories. Hence, we conjecture that if the UV completion of such a theory exists, it must be a 6d $\cN=(1,0)$ SCFT.
	\item The number of tensor multiplets required to cancel the gauge anomaly (via Green Schwarz mechanism) is one less than the number of gauge groups. There is a BPS instanton string in the theory whose tension is not controlled by the vevs of the scalars in the tensor multiplets. This tension becomes a mass parameter of the theory. We conjecture that if the UV completion of such a theory exists, it must be a 6d $\cN=(1,0)$ little string theory.
\end{itemize}

While this paper was near its completion, we received \cite{Heckman:2015bfa} which has a significant amount of overlap with this work.

The paper is organized as follows. In section \ref{2}, we review certain aspects of 6d $\cN=(1,0)$ gauge theories relevant for the study of this work. In section \ref{3}, we describe the necessary consistency conditions which we demand from gauge theories in our classification. In section \ref{4}, we delineate the classification procedure. We carry out the procedure fully for the case in which all the simple gauge group factors are classical groups and the matter is in half/full hypermultiplets transforming in bifundamentals and fundamentals of the simple gauge factors. In section \ref{5}, we present a summary of the work and comment on some possible future directions.

\section{Review of 6d $\cN=(1,0)$ gauge theories} \label{2}
We start by reviewing some aspects of 6d $\cN=(1,0)$ gauge theories relevant for the study of this work. A large part of this section is taken from \cite{Seiberg:1996qx}.

6d $\cN=(1,0)$ theories have 8 real supercharges. These theories admit the following field multiplets
\begin{enumerate}
	\item \textbf{Vector multiplet:} The bosonic field content is only a vector field $A$. Hence, these theories do not have a Coulomb branch of vacua.
	\item \textbf{Hypermultiplet:} The bosonic field content is 4 real scalars. These scalars parametrize the Higgs branch of vacua.
	\item \textbf{Tensor multiplet:} The bosonic field content is a two-form $B$ and a real scalar $\phi$. The field strength $H$ corresponding to $B$ is constrained to be self dual, that is $H=\star H$. The real scalar $\phi$ parametrizes the tensor branch of vacua.
\end{enumerate}

For the purpose of illustration, we consider only the case of a simple gauge group in this section. First consider a consistent gauge theory with matter only in hypermultiplets. Then, the gauge coupling is a dimensionful parameter of the theory. This follows from the fact that the gauge field $A$ must have mass dimension 1 for it to mix with ordinary derivative to form covariant derivative. Thus, for a typical process with characteristic energy scale $E$, the strength of the interactions would be captured by the dimensionless quantity $s=g^2E^2$. We see that such a theory is trivial in the IR but lacks a proper definition in the UV. The theory has instantons which, because we are in 6d, are string like excitations. Notice that the gauge kinetic term
\begin{equation}
\frac{1}{g^2}\int \text{ tr}(F\wedge\star F)
\end{equation}
implies that a BPS instanton satisfying $F=\star F$ has a tension
\begin{equation}
T_{BPS}\propto\frac{1}{g^2}
\end{equation}
Since the tension of these strings is a parameter of the theory, it seems that the winding modes of these strings might give rise to a notion of T-duality in this theory. This suggests that there might be a good chance that this theory has a UV completion in the form of a theory which enjoys T-duality. In fact, using string theory arguments, many such gauge theories have been found as the low energy limits of little string theories which are UV complete theories having T-duality \cite{Seiberg:1997zk,Intriligator:1997dh}.

Now consider a consistent gauge theory with matter in hypermultiplets and a single tensor multiplet. Because the theory contains a self-dual field strength $H$, it is not known how to write down a useful Lagrangian for the theory. In what follows, we will pretend that the interactions are described by conventional Lagrangian interaction terms with the added constraint of $H=\star H$. These interaction terms should legitimately be viewed as a convenient representation of equations of motion (which are well-defined). The tensor multiplet scalar couples to the gauge field with an interaction term of the form
\begin{equation}
c\int\phi\text{ tr}(F\wedge\star F) \label{interaction}
\end{equation} where $c$ is a constant. We absorb the gauge coupling into $\phi$ by a field redefinition. On the tensor branch, this term gives rise to an effective gauge coupling for the theory
\begin{equation}
\frac{1}{g^2_{eff}}=c\langle\phi\rangle
\end{equation}
Thus, we see that the effective gauge coupling of the theory is controlled by the tensor modulus. A BPS instanton now has a tension which is controlled by the vev of scalar in the tensor multiplet
\begin{equation}
T_{BPS}\propto\frac{1}{g_{eff}^2}=c \langle\phi\rangle
\end{equation}
Hence, the tension of these strings is not a parameter of the theory and winding modes cannot give rise to a notion of T-duality. But, instead, one can notice that the only dimensionful quantity $T_{BPS}$ vanishes at the origin of tensor branch hinting at the possibility of a strongly coupled fixed point. This suggests that there might be a good chance that this theory has a UV completion in the form of a theory enjoying scale invariance. In fact, using string theory arguments, many such gauge theories have been found as the low energy limits on tensor branch of SCFTs \cite{Hanany:1997gh}.

(\ref{interaction}) and supersymmetry imply that there must exist an interaction term of the form
\begin{equation}
c\int B\wedge\text{tr}(F\wedge F)
\end{equation}
which is equivalent to the equation of motion
\begin{equation}
d\star H=c\text{ tr}(F\wedge F)
\end{equation}
Self-duality then implies that the Bianchi identity of $H$ is modified to
\begin{equation}
dH=c\text{ tr}(F\wedge F)
\end{equation}
This is precisely the statement that instantons are charged under the two-form $B$. These charges must be quantized by an analog of the Dirac quantization condition. This means that $c^2$ must be quantized. Also, via descent equations, the modification of Bianchi identity provides a contribution to the anomaly polynomial of the form
\begin{equation}
I_8=-c^2(\text{tr}(F\wedge F))^2
\end{equation}
Hence, in a consistent gauge theory, the coefficient of the quadratic part of the anomaly polynomial coming from one-loop gauge anomaly must be $c^2\ge0$. In other words, even if the quadratic part of the 1-loop anomaly polynomial does not vanish, the gauge theory can be made consistent by the addition of a tensor multiplet as long as the coefficient of quadratic part is positive definite. This is an illustration of Green-Schwarz mechanism of anomaly cancellation in 6d gauge theories. Also notice that the above mentioned quantization of $c^2$ requires that the coefficient of the quadratic part of 1-loop anomaly polynomial be appropriately quantized.

\section{Consistency conditions on gauge theories} \label{3}
\subsection{Anomaly cancellation}
Consider a 6d $\cN=(1,0)$ gauge theory with the gauge group
\begin{equation}
G=G_1\times G_2\times \cdots G_s,
\end{equation}
(where $G_a$ is simple) along with full/half hypermultiplets in the representations $R_p$ where
\begin{equation}
R_p=R_{1,p}\otimes R_{2,p} \otimes \cdots \otimes R_{s,p}
\end{equation}
with $R_{a,p}$ being an irreducible complex representation of $G_a$. $
R_{a,p}$ can be a trivial one-dimensional representation.
The pure gauge part of the anomaly polynomial can be written as
\begin{equation}
I_8=\sum_a \text{Tr}F_a^4-\sum_p\eta_p\text{Tr}_{R_p}(\sum_a1\otimes\cdots\otimes1\otimes F_a\otimes1\otimes\cdots\otimes1)^4
\end{equation}
where $\eta_p=1/2$ if there is a half-hypermultiplet in representation $R_p$ and $\eta_p=1$ if there is a full hypermultiplet in representation $R_p$. Throughout this paper, Tr denotes a trace in the adjoint representation and Tr$_R$ denotes a trace in the representation $R$. Except for the case of $\SO(8)$, the above can always be rewritten as 
\begin{equation}
I_8=\alpha^a\text{tr}F_a^4+c^{ab}\,(\text{tr}F_a^2)(\text{tr}F_b^2)
\end{equation}
where $c^{ab}=c^{ba}$. Here tr denotes a trace in a representation of our choice for each group. We choose this to be the fundamental representation for $\SU(n)$ and $\USp(n)$; the vector for $\SO(n\ge7)$, $\mathsf{E}_6$, $\mathsf{E}_7$, $\mathsf{F}_4$ and $\mathsf{G}_2$; and the adjoint for $\mathsf{E}_8$. We will call this chosen representation the \emph{fundamental} representation of the group by a slight abuse of terminology. $\SO(8)$ has three independent Casimir operators of degree 4 \cite{Grassi:2011hq}, which can be taken to be tr$F^4$, Tr$_{\spinor}F^4$ and Tr$_{\conj}F^4$ where $\spinor$ and $\conj$ are two spinor representations of $\SO(8)$ and
\begin{equation}
(\text{tr}F^2)^2=\frac{1}{3}\left(\text{tr}F^4+\text{Tr}_\spinor F^4+\text{Tr}_\conj F^4\right)
\end{equation}
From now on, our discussion will assume generically that the gauge group is not $\SO(8)$. The case of $\SO(8)$ will be discussed separately.

The gauge part of the anomaly polynomial must vanish for the theory to make sense on $\mathbb{R}^6$. This requires that at least 
\begin{equation}
\alpha^a=0\:\:\:\:\:\:\forall a
\end{equation}
Now, suppose that the rest of the anomaly factorizes as 
\begin{equation}
c^{ab}\,(\text{tr}F_a^2)(\text{tr}F_b^2)=\sum_i(k_i^a\text{ tr}F_a^2)^2
\end{equation}
Then, this can be cancelled by adding a number of $(1,0)$ tensor multiplets equal to the number of terms in the above sum. Specifically, one modifies the Bianchi identity of the $i$th self dual field strength to 
\begin{equation}
dH_i=k_i^a\text{ tr}F_a^2
\end{equation}
The tensor multiplet $i$ then contributes a term to the anomaly polynomial equal to $-\sum_i(k_i^a$ tr$F_a^2)^2$ which cancels against the quadratic part mentioned above. This is known as the Green-Schwarz mechanism of anomaly cancellation and is explained well in \cite{Brunner:1998vfa}. This is equivalent to the condition that the matrix $c^{ab}$ is positive semi-definite. When it is positive definite, the number of tensor multiplets required to cancel the anomaly is equal to the number of gauge groups and hence the gauge theory should have a UV completion (if it exists) in the form of a $(1,0)$ SCFT. When it is not positive definite but positive semi-definite with one zero eigenvalue, then we conjecture it to be a little string theory based on the reason mentioned in the introduction. Other cases with multiple zero eigenvalues do not appear in our classification.

Now, consider a $(1,0)$ gauge theory which might have a UV completion in the form of an SCFT. One would like to be able to put a CFT on an arbitrary background. For the theory to make sense on an arbitrary background, the mixed gauge-gravitational part of the anomaly polynomial must also vanish. It turns out that, in the case when the number of tensor multiplets equals the number of gauge groups, one can always modify the Green-Schwarz mechanism without the addition of any new tensor multiplets to cancel the mixed anomaly too. This modification does not modify the anomaly cancellation conditions written above \cite{Ohmori:2014kda}.

In the case of $\SO(8)$,
\begin{equation}
\text{Tr}F^4=4\left(\text{tr}F^4+\text{Tr}_\spinor F^4+\text{Tr}_\conj F^4\right)=12(\text{tr}F^2)^2
\end{equation}
and hence the Green-Schwarz mechanism can be used to cancel the anomaly only if the number of vectors, spinors and conjugate spinors are equal in number. Here we are assuming that the higher dimensional irreps contribute such that Green-Schwarz anomaly cancellation is not possible. See section \ref{assumption} for an analogous assumption in the case of $G\ne\SO(8)$.

\subsection{Global anomaly}\label{global}
There is also a global anomaly which affects $\SU(2)$, $\SU(3)$ and $\mathsf{G}_2$ respectively \cite{Bershadsky:1997sb}:
\begin{align}
4-n_2&=0\:\:\:\text{mod }6\\
n_3-n_6&=0\:\:\:\text{mod }6\\
1-n_7&=0\:\:\:\text{mod }3
\end{align}
where $n_2$ is the number of full hypers in the doublet of $\SU(2)$; $n_3$ and $n_6$ are the number of hypers in the fundamental and symmetric representations of $\SU(3)$ respectively; and $n_7$ is the number of hypers in the fundamental of $\mathsf{G}_2$. Here $n_2$ also includes a contribution from a full hyper charged in fundamental of $\SU(2)$ and some representation $R$ of another group $G$. The contribution is equal to dimension of $R$. There are similar contributions from representations charged under $\SU(2)$ and two or more other groups. The same is true for $n_3$, $n_6$ and $n_7$ where instead we look at representaions charged under $\SU(3)$, $\SU(3)$ and $\mathsf{G}_2$ respectively. In addition to above contributions, $n_7$ also receives contributions from a half hyper charged in the fundamental of $\mathsf{G}_2$ and some pseudo-real representation $R$ of another group $G$. This contribution is equal to half the dimension of $R$. 

\subsection{Quantization of charges of instanton strings}
Incidentally, there is an extra consistency check one can perform on the theories satisfying above mentioned conditions. In the Green-Schwarz mechanism recalled above, the Bianchi identity for $H_i$ was modified. It is clear from the modification that instanton configurations of gauge fields are charged under B and $k_i^a$ are related to the charges of these instantons. In 6d, instantons are string like excitations in the theory and hence their charges must be appropriately quantized. The fact that the term on the right hand side of the modified Bianchi identity also appears in the Green-Schwarz contribution to the anomaly means that the matrix $c^{ab}$ must also be appropriately quantized. The full details of the argument can be found in \cite{Ohmori:2014kda}. This condition translated into our notation is
\begin{equation}
M^{ab}=\frac{c^{ab}n^an^b}{12}\in \mathbb{Z}\:\:\: \forall\, a,b
\end{equation}
where $n^a$ is an integer assigned to every group and is listed in Table~\ref{ni}.
\begin{table}
\centering
\begin{tabular}{c| c }
Group & $n^a$ \\
\hline
\hline
$\SU(n)$ & 2 \\
$\SO(n)$ & 4 \\
$\USp(n)$ & 2 \\
$\mathsf{G}_2$ & 4 \\
$\mathsf{F}_4$ & 12 \\
$\mathsf{E}_6$ & 12 \\
$\mathsf{E}_7$ & 24 \\
$\mathsf{E}_8$ & 120 \\
\end{tabular}
\caption{List of integers relevant for the quantization condition\label{ni}}
\end{table}

Incidentally, the matrix $M$ can be thought of as controlling the kinetic term for the scalars in the tensor multiplets (where we add a few decoupled tensor multiplets corresponding to zero eigenvalues of $M$). This requires that $M$ must be positive semi-definite but this is not a new consistency condition because the positive semi-definiteness of $M$ is equivalent to the positive semi-definiteness of the matrix $c^{ab}$. To see this, recall the fact that a matrix is postive semi-definite if and only if the determinant of every principal sub-matrix (including the full matrix itself) is non-negative. The determinant of a principal sub-matrix of $M$ is equal to the determinant of the corresponding principal submatrix of the matrix $c^{ab}$ times $\prod_p\frac{n^p}{\sqrt{12}}\prod_q\frac{n^q}{\sqrt{12}}$ where $p$ runs over the rows and $q$ runs over the columns of the principal sub-matrix. As all $n^a$ are positive numbers, our above claim is justified and we need only demand the positive semi-definiteness of the matrix $c^{ab}$.

\section{Classification} \label{4}
\subsection{An assumed restriction on the allowed representations} \label{assumption}
For any representation $R$ of a simple group $G$ we define $\alpha_R$ and $c_R$ through
\begin{equation}
\text{Tr}_R F^4=\alpha_R \text{ tr}F^4+c_R (\text{tr}F^2)^2
\end{equation}
The condition that the matrix $c^{ab}$ is positive semi-definite is equivalent to the condition that the determinant of every principal submatrix (including the full matrix itself) is non-negative. In particular, this means that every diagonal entry $c^{aa}$ is non-negative
\begin{equation}
c^{aa}=c_{adj}-\sum_{p}\eta_p c_{R_{a,p}}\prod_{b\neq a} d_{R_{b,p}}\ge0
\end{equation}
where $d_R$ denotes the dimension of the representation $R$. We restrict our analysis to those irreps $R_{a,p}$ that satisfy 
\begin{itemize}
\item $d_{R_{a,p}}\le d_{adj_a}$ for a complex or strictly real $R_{a,p}$.
\item $d_{R_{a,p}}\le 2\,d_{adj_a}$ for a pseudo-real $R_{a,p}$.
\end{itemize}
All the irreps satisfying these conditions have $c_{R_{a,p}}>0$. This means that we need to only look at irreducible representations $R_{a,p}$ of $G_a$ such that 
\begin{equation}
\eta_p c_{R_{a,p}}\le c_{adj_a} \label{c}
\end{equation}
We are not aware of any consistent $(1,0)$ gauge theory which we would not be able to see because of this restriction. The irreps satisfying (\ref{c}) are listed in Table~\ref{data-infinite} and Table~\ref{data-isolated} \cite{Erler:1993zy}. We don't mention the symmetric traceless irrep of $\SO(n)$ because if there is a hyper in this irrep then it is not possible to obtain $\alpha_{\SO(n)}=0$.

\begin{table}
\centering
\begin{tabular}{c |c | c | c | c }
Type & Name  & dimension  &  $\alpha_R$   & $c_R$ \\
\hline
\hline
\multicolumn{5}{c}{$\SU(n\ge4)$} \\
\hline
Complex &   $\fund$ & $n$ & $1$ & $0$ \\
Complex &   $\asym$ & $\frac{n(n-1)}{2}$ & $n-8$ & 3 \\
Complex &  $\sym$ & $\frac{n(n+1)}{2}$ & $n+8$ & $3$ \\
Strictly real &   $\adj$ & $n^2-1$ & $2n$ & $6$ \\
\hline
\multicolumn{5}{c}{$\SO(n\ge7,n\ne8)$} \\
\hline
Strictly real &  $\vect$ & $n$ & $1$ & $0$  \\
Strictly real &  $\adj $ & $n(n-1)/2$ & $n-8$ & 3 \\
\hline
\multicolumn{5}{c}{$\USp(n\ge4)$} \\
\hline
Pseudo-real &  $\vect$ & $n$ & 1 &  $0$ \\
Strictly real &  $\asym$  & $\frac{(n+1)(n-2)}{2}$ & $n-8$ & 3 \\
Strictly real &  $\adj$ & $\frac{n(n+1)}{2}$ & $n+8$ & 3 \\
\hline
\end{tabular}
\caption{List of allowed representations for single gauge group: infinite series.
$\fund$ : fundamental, $\asym$: two-index antisymmetric tensor, $\sym$: two-index symmetric tensor, $\adj$: adjoint, $\vect$: vector. The $\asym$ for $\USp$ is the antisymmetric traceless representation.
We also sometimes call $\vect$ of $\SO$ and $\USp$ as $\fund$, if no confusion arises. \label{data-infinite}}
\end{table}

\begin{table}
\centering
\begin{tabular}{c| c  | c | c | c |c }
Group & Type & Name & Dimension  &  $\alpha_R$   & $c_R$  \\
\hline
\hline
$\SU(2)$ & Pseudo-real &  $\fund$ & 2 & 0 &  1/2 \\
$\SU(2)$ & Strictly real &  $\adj$ & 3 & 0 &  8 \\
$\SU(3)$ & Complex &  $\fund$ & 3 & 0 &  1/2 \\
$\SU(3)$ & Complex &  $\sym$ & 6 & 0 &  17/2 \\
$\SU(3)$ & Strictly real  & $\adj$ & 8 & 0 &  9 \\
$\SU(6)$ & Pseudo-real &   $\asymT$ & 20 & $-6$ &  6 \\
\hline
$\SO(7)$ & Strictly real  &  $\spinor$ & $8$ & $-1/2$ &  3/8 \\
$\SO(9)$ & Strictly real  &  $ \spinor$ & $16$ & $-1$ &   3/4\\
$\SO(10)$ & Complex    &    $\spinor$ & $16$ & $-1$ & 3/4  \\
$\SO(11)$ & Pseudo-real   &   $\spinor$ & $32$ & $-2$ &  3/2 \\
$\SO(12)$ & Pseudo-real   &   $\spinor$ & $32$ & $-2$ &  3/2 \\
$\SO(13)$ & Pseudo-real   &   $\spinor$ & $64$ & $-4$ &  3 \\
$\SO(14)$ & Complex  &  $\spinor$ & $64$ & $-4$ &  3 \\
\hline
$\USp(6)$ & Pseudo-real  &  $\asymT$ & $14$ & $-7$ &  6 \\
\hline
$\mathsf{E}_6$ & Complex   &$\fund$& $27$ & $0$ & 1/12 \\
$\mathsf{E}_6$ & Strictly real  &$\adj$& $78$ & $0$ & 1/2 \\
$\mathsf{E}_7$ & Pseudo-real   &$\fund$& $28$ & $0$ & 1/24 \\
$\mathsf{E}_7$ & Strictly real  &$\adj$& $133$ & $0$ & 1/6 \\
$\mathsf{E}_8$ & Strictly real  &$\adj$& $248$ & $0$ & 1/100 \\
$\mathsf{F}_4$ & Strictly real  &$\fund$& $26$ & $0$ & 1/12 \\
$\mathsf{F}_4$ & Strictly real  &$\adj$& $52$ & $0$ & 5/12 \\
$\mathsf{G}_2$ & Strictly real &$\fund$& $7$ & $0$ & 1/4 \\
$\mathsf{G}_2$ & Strictly real  &$\adj$& $14$ & $0$ & 5/2 \\
\end{tabular}
\caption{List of allowed representations for single gauge group: isolated ones.
$\asymT$: three-index antisymmetric,
$\spinor$: spinor representation.
We don't distinguish between the two spinors of $\SO(12)$ in our classification. 
\label{data-isolated}}
\end{table}

\subsection{Some illustrations}
We think that at this point we should illustrate how some of the theories already known in the literature arise in our analysis.

For instance, it is known that $\SO(7)$ gauge theory with two full hypers in the spinor representation $\spinor$ is consistent \cite{Sadov:1996zm}. Let's see why this is so in our notation:
\begin{align}
	I_8=\text{Tr}F^4-2\text{ tr}_\spinor F^4 \label{ill1}
\end{align}
Using Table~\ref{data-infinite} we see that
\begin{align}
	\text{Tr}F^4=-\text{tr}F^4+3(\text{tr}F^2)^2
\end{align}
and using Table~\ref{data-isolated} we see that
\begin{align}
	\text{tr}_\spinor F^4=-\frac{1}{2}\text{tr}F^4+\frac{3}{8}(\text{tr}F^2)^2
\end{align}
So, (\ref{ill1}) becomes
\begin{align}
	I_8=\frac{9}{4}(\text{tr}F^2)^2
\end{align}
Thus, $\alpha=0$ and $c=9/4$, and the anomaly can be cancelled by adding a single tensor multiplet. Let's compute the matrix controlling the kinetic term for scalar in the tensor multiplet
\begin{align}
	M=\frac{c n^2}{12}
\end{align}
and reading $n$ from Table~\ref{ni} we find that
\begin{align}
	M=3
\end{align}
which is an integer, as required by the quantization condition. Hence, we conclude that this theory is consistent.

A well known example is $\SU(n)\times\SU(n)$ gauge theory with a full hyper in the bifundamental and $n$ full hypers in the fundamental representation of each gauge group. In this case
\begin{align}
	I_8=&\text{Tr}F_1^4+\text{Tr}F_2^4-n\text{ tr}F_1^4-n\text{ tr}F_2^4-\text{tr}_{\fund\otimes\fund}(F_1\otimes1+1\otimes F_2)^4\\
	=&2n\text{ tr}F_1^4+6(\text{tr}F_1^2)^2+2n\text{ tr}F_2^4+6(\text{tr}F_2^2)^2-n\text{ tr}F_1^4-n\text{ tr}F_2^4 \nonumber \\
	&-\text{tr}F_1^4d_\fund-d_\fund\text{tr}F_2^4-6\text{ tr}F_1^2\text{ tr}F_2^2\\
	=&6(\text{tr}F_1^2)^2+6\text{tr}(F_2^2)^2-6\text{ tr}F_1^2\text{ tr}F_2^2
\end{align}
Hence, we see that $\alpha_1=\alpha_2=0$, $c_{11}=c_{22}=6$ and $c_{12}=c_{21}=-3$. Thus, the anomaly can be cancelled by adding two tensor multiplets. We can calculate the matrix controlling the kinetic terms of tensor multiplet scalars yielding $M_{11}=M_{22}=2$ and $M_{12}=M_{21}=-1$ which satisfies the quantization condition.

\subsection{Theories with simple gauge group}
The classification for the case of simple  gauge group was already done in \cite{Danielsson:1997kt}. Here we re-derive this as a sub-result of our classification.

From now on we only consider theories such that for every $G_a$ there is at least one hyper charged non-trivially under $G_a$ and some other group $G_b$. In a sense, to be made precise in the form of quiver diagrams later, we are only looking at \emph{connected} theories. Any other theory can be seen as a disjoint union of connected theories. Hence, we only need to classify the connected ones.

An off diagonal element of the matrix $[c^{ab}]$ can be written as
\begin{equation}
c^{ab}=-\frac{1}{2}\sum_{p}\eta_p C_p i_{R_{a,p}}i_{R_{b,p}}\prod_{c\neq a,b} d_{R_{c,p}}
\end{equation}
where $C_p$ is a combinatorial factor and the index $i_{R_{a,p}}$ is defined by
\begin{equation}
\text{Tr}_{R_{a,p}}F_a^2=i_{R_{a,p}}\text{tr}F_a^2
\end{equation}
Therefore,
\begin{equation}
c^{ab}<0\:\:\:\:\:\:\forall a\neq b
\end{equation}
This means that if $c^{aa}=0$ for some $G_a$, then the full gauge group of the theory must be $G=G_a$ and the theory would be a potential little string theory. We can show it by contradiction. Suppose there is another simple factor $G_b$ in $G$. Then, the determinant of the two by two principal submatrix formed by row $a$ and row $b$ would be negative.

Hence, we are already done with the classification of the potential little string theories based on simple gauge group. These are the theories with $c^{aa}=0$
\begin{itemize}
	\item Any group with 1 hyper in $\adj$.
\item $\SU(n\ge4)$ with 2 hypers in $\asym$ and 16 hypers in $\fund$. $\SU(3)$ with 18 hypers in $\fund$. $\SU(2)$ with 16 full hypers in $\fund$.
\item $\SU(n\ge4)$ with 1 hyper in $\asym$ and 1 hyper in $\sym$. $\SU(3)$ with 1 hyper in $\fund$ and 1 hyper in $\sym$.
\item $\SU(6)$ with 1 full hyper in $\asymT$ and 18 hypers in $\fund$.
\item $\SU(6)$ with 1 half-hyper in $\asymT$, 1 hyper in $\fund$ and 1 hyper in $\sym$.
\item $\SO(7\le n\le14, n\ne8)$ with $2^{7-\lfloor(n+1)/2\rfloor}$ full hypers in $\spinor$ and $n-4$ hypers in $\fund$. Here $\lfloor r\rfloor$ denotes the greatest integer less than or equal to $r$.
\item $\SO(8)$ with 4 hypers each in $\vect$, $\spinor$ and $\conj$.
\item $\USp(2n\ge4)$ with 1 hyper in $\asym$ and 16 hypers in $\fund$.
\item $\USp(6)$ with 1 half-hyper in $\asymT$ and 35 half-hypers in $\fund$.
\item $G_2$ with 10 hypers in 7 dimensional rep.
\item $F_4$ with 5 hypers in 26 dimensional rep.
\item $E_6$ with 6 hypers in 27 dimensional rep.
\item $E_7$ with 4 hypers in 56 dimensional rep.
\end{itemize}

We will find that most of the potential SCFTs with simple gauge group arise when we take the quiver size to be 1 of the generalized quiver gauge theories that we will introduce later. So, we don't mention all those potential SCFTs having simple gauge group here. However, we will soon see that the exceptional groups except $\mathsf{G}_2$ don't couple to any other group and have $\alpha=0$. So, we write potential SCFTs having simple exceptional group here:
\begin{itemize}
	\item $\mathsf{E}_6$ with less than or equal to 5 hypers in $\fund$.
	\item $\mathsf{E}_7$ with less than or equal to 7 half-hypers in $\fund$.
	\item $\mathsf{F}_4$ with less than or equal to 4 hypers in $\fund$.
	\item $\mathsf{G}_2$ with $1,4,7$ hypers in $\fund$. These are the only allowed theories because of the global anomaly mentioned in section \ref{global}
\end{itemize}

For now, one can manually check that the quantization condition is satisfied for all the above mentioned theories. Later we will give a proof that any theory satisfying the constraints of anomaly cancellation will automatically satisfy the quantization condition too.

\subsection{Possible hypers between multiple groups}

\begin{table}
\centering
\begin{tabular}{r@{$\times$}l|c|c@{$\otimes$}c}
$G_a$  & $G_b$ & half hyper? & $R_a$ & $R_b$\\
\hline
\hline
$\SU(n)$ & $\SU(m)$ & full & $\fund$ & $\fund$\\
$\SO(n)$ & $\USp(m)$ &  half & $\fund$ & $\fund$\\
$\SU(m)$ & $\SO(n)$ & full &$ \fund$ & $\fund$\\
$\SU(m)$ & $\USp(n)$ & full & $\fund$ & $\fund$\\
$\USp(m)$ & $\USp(n)$ & full & $\fund$ & $\fund$\\
\hline
$\SO(7,8)$ & $\SU(4)$ & full & $\spinor$ & $\fund$\\
$\mathsf{G}_2$ & $\SU(4)$ & full & $\fund$ & $\fund$\\
$\SU(4)$ & $\SU(2)$ & half & $\asym$ & $\fund$\\
$\SO(7,8)$ & $\USp(n\le12)$ & half & $\spinor$ & $\fund$\\
$\mathsf{G}_2$ & $\USp(n\le14)$ & half & $\fund$ & $\fund$\\
$\SO(7,8)$ & $\SU(2)$ & half & $\spinor$ & $\fund$\\
$\mathsf{G}_2$ & $\SU(2)$ & half & $\fund$ & $\fund$\\
$\SO(7,8)$ & $\SU(2)$ & half & $\fund$ & $\fund$
\end{tabular}
\caption{List of hypers for $G_1\times G_2$.
 \label{2-gon}}
\end{table}

There are a few possible hypers charged under two groups and they are collected in Table~\ref{2-gon}. Some of these combinations already have determinant zero. Let's say we couple another group to such \emph{determinant 0} combinations by adding new hypers charged under the new group and old groups. Then evaluate the determinant around new row. We will get a sum of $2\times2$ determinants with one of them being the determinant of the submatrix corresponding to the old groups. After the addition of new matter, the diagonal entries of this submatrix either decrease or stay the same and the off diagonal entries remain the same. Hence this determinant is non-positive. The other determinants can be expanded once again, but this time we expand them around the new column to yield $1\times1$ determinants. It is easy to see that, for every term, the total coefficient (coming from expansion) times the $1\times1$ determinant is strictly negative. Hence, it is not possible to couple a determinant 0 combination of two groups to other groups and matter. Using this and similar computations, one can show that it is not possible to couple other groups and matter to a determinant zero combination of $n$ groups. So, we obtain a few more potential little string theories:
\begin{itemize}
	\item $\USp(m)\times\USp(n)$ theory with a hyper in $\fund\otimes\fund$, $m+8-n$ full hypers in $\fund$ of $\USp(m)$ and $n+8-m$ full hypers in $\fund$ of $\USp(n)$. Of course, $|m-n|\le8$ for the theory to exist.
	\item $\SO(7)\times\SU(4)$ theory with a hyper in $\spinor\otimes\fund$ and $1$ hyper in $\fund$ of $\SO(7)$.
	\item $\mathsf{G}_2\times\SU(4)$ theory with a hyper in $\fund\otimes\fund$ and $1$ hyper in $\fund$ of $\SU(4)$
\end{itemize}

There is only one possible hyper charged under three groups and it gives rise to a potential little string theory
\begin{itemize}
	\item $\SU(2)\times\SU(2)\times\SU(2)$ theory with a half hyper in $\fund\otimes\fund\otimes\fund$ and a full hyper in $\fund$ of each $\SU(2)$.
\end{itemize}
There are no possible hypers charged under more than three groups.

We now turn to the proof that quantization condition is automatically satisfied for all the theories satisfying constraints of anomaly cancellation. The proof is just based on a few observations and we don't have a deep reason why every \emph{anomaly cancellable} theory must satisfy quantization condition.

\subsection{Quantization condition is automatically satisfied}
Consider a full/half hypermultiplet charged in a representation $R$ of a gauge group $G$. From Table~\ref{data-infinite} and Table~\ref{data-isolated}, we see that $c_R\times n_G^2/12$ is an integer for all cases except $\spinor$ of $\SO(7)$ for which it is half-integer.

Next, consider a full/half hypermultiplet charged in $R_1\otimes R_2$ of $G_1\times G_2$. From Table~\ref{2-gon}, we see that all contributions to the matrix are integers for all cases except $\half\spinor\otimes\fund$ of $\SO(7)\times\USp(2,6,10)$ for which the contribution to the diagonal entry corresponding to $\SO(7)$ is half-integral. We write $\SU(2)$ as $\USp(2)$ for this subsection only.

Now, notice that for both of these exceptions, their contribution to $\alpha_{\SO(7)}$ is half-integral but for all other combination of representations the contribution is integral. Hence, if we want to arrange a configuration such that $\alpha_{\SO(7)}=0$, we must include these exceptions even number of times. But this means that their combined contribution to the diagonal entry corresponding to $\SO(7)$ is integral too.

For the trifundamental, it can be manually checked that the quantization condition is satisfied. This completes the proof that if a theory satisfies the constraints of anomaly cancellation, then it satisfies quantization condition too. Combining this result with our previous result that the positive semi-definiteness of $M$ is equivalent to the positive semi-definiteness of the matrix $c^{ab}$, we see that the matrix $M$ provides no new consistency conditions over the anomaly cancellation constraints. Thus, we will not be computing the matrix $M$ from now on.

\subsection{Quivers and branches}
Notice from Table~\ref{2-gon} that the majority of combinations are bifundamentals between classical gauge groups. We first consider all theories with classical gauge groups and full/half hypers in bifundamental and fundamental representations. For such theories, we introduce quiver diagrams
\begin{itemize}
	\item \begin{tikzpicture}
\node[SU] (A) at (0,0) {$n$};
\end{tikzpicture}
 denotes an $\SU(n)$ group, 
\begin{tikzpicture}
\node[Sp] (A) at (0,0) {$2n$} ;
\end{tikzpicture}
denotes a $\USp(2n)$ group, and
\begin{tikzpicture}
\node[SO] (A) at (0,0) {$n$} ;
\end{tikzpicture}
denotes an $\SO(n)$ group. We also allow nodes of the form \begin{tikzpicture}
\node[SU] (A) at (0,0) {$2$};
\end{tikzpicture}, \begin{tikzpicture}
\node[SU] (A) at (0,0) {$3$};
\end{tikzpicture} and \begin{tikzpicture}
\node[Sp] (A) at (0,0) {$2$};
\end{tikzpicture}. At this point these are formal nodes since these gauge groups behave differently than the other ones in the same classical family. Later, we will give a meaning to these nodes.
\item An edge between $\SU(n)$ and any other group $X$ denotes a full hyper in $\fund\otimes\fund$ of $\SU(n)\times X$. An edge between $\SO(m)$ and $\USp(n)$ denotes a half hyper in $\fund\otimes\fund$ of $\SO(m)\times\USp(n)$. We don't have to consider edges between two $\USp$ and two $\SO$ nodes because in the former case only one theory is allowed which we have already listed and in the latter case there is no allowed theory.
\end{itemize}

As all the traces involved are in fundamentals and bifundamentals, the diagonal entries of $[c^{ab}]$ are independent of the matter content of the theory and depend only on the type of group involved. It is 6 for $\SU$ and 3 for $\SO$ and $\USp$. This means that if a diagram is allowed then any subdiagram formed by a subset of nodes and all the edges between these subset of nodes is also allowed. Therefore, we can construct all allowed theories by adding one node (with any number of edges) at a time to an allowed theory and checking if the resulting theory is allowed or not. This is captured completely by the determinant of $[c^{ab}]$. An allowed theory has non-negative determinant. If the determinant is 0, adding any new node will make the determinant negative. The original allowed theory is a potential little string theory. If the determinant is positive, then one has to check the determinant of the resulting theory after adding the new node. 

The above procedure can be carried out by first forgetting about the labels of the nodes and just classifying the structure of the diagrams. Then, one can put the labels in the nodes and see if there exists some consistent set of labels for every structure. To list all the possible labellings, we find it convenient to introduce some more terminology
\begin{itemize}
	\item When there is single edge between $\SU(m)$ and $X(n)=\SU(n), \USp(n), \SO(n)$, we say there is a current of $m-n$ from $\SU(m)$ to $X(n)$ and a current of $n-m$ from $X(n)$ to $\SU(m)$.
	\item When there is a single edge between $\USp(m)$ and $\SO(n)$, we say there is a current of $m+8-n$ from $\USp(m)$ to $\SO(n)$ and a current of $n-8-m$ from $\SO(n)$ to $\USp(m)$.
	\item Sometimes we denote the current $i$ from $X(m)$ to $Y(n)$ as a directed edge from the node $X(m)$ to node $Y(n)$ with $i$ written on top of this directed edge. Notice that the direction of the arrow should not be confused with the direction of flow of positive current. There can be a directed edge with $i<0$.
\end{itemize}
We define three types of \emph{branches} in the spirit of \cite{Bhardwaj:2013qia}:
\paragraph{$\SU(n_0)$ branches:}
Such a branch is composed of a chain of $\SU$ nodes starting with a node $\SU(n_0)$ such that every node except $\SU(n_0)$ has $\alpha=0$. We often suppress the square nodes corresponding to hypers in fundamental of a node while writing a branch as they can always be figured out from the currents in the edges emanating from the node. Let's call the node following $\SU(n_0)$ as $\SU(n_1)$. We define the branch current as $n_0-n_1$. Notice that the current in a particular direction is monotonically increasing.

\paragraph{$\SO(n_0)$ branches:}
Such a branch is composed of a chain of alternating $\SO$-$\USp$ nodes starting with a node $\SO(n_0)$ such that every node except $\SO(n_0)$ has $\alpha=0$. We often suppress the square nodes corresponding to hypers in fundamental. Let's call the node following $\SO(n_0)$ as $\USp(n_1)$. We define the branch current as $n_0-8-n_1$. Notice that the current in a particular direction is monotonically increasing.

\paragraph{$\USp(n_0)$ branches:}
Such a branch is composed of a chain of alternating $\USp$-$\SO$ nodes starting with a node $\USp(n_0)$ such that every node except $\USp(n_0)$ has $\alpha=0$. We often suppress the square nodes corresponding to hypers in fundamental. Let's call the node following $\USp(n_0)$ as $\SO(n_1)$. We define the branch current as $n_0+8-n_1$. Notice that the current in a particular direction is monotonically increasing.

\subsection{Classification of bifundamentals and fundamentals}
From now on, we often suppress hypers in $\fund$.

Carrying out the classification process outlined above, we obtain the following list of allowed theories composed only of bifundamentals and fundamentals.

\paragraph{Potential little string theories:}
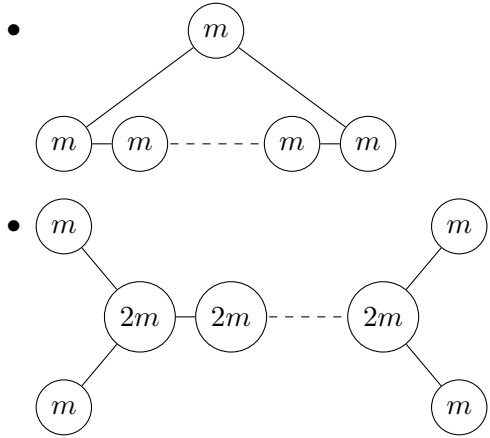
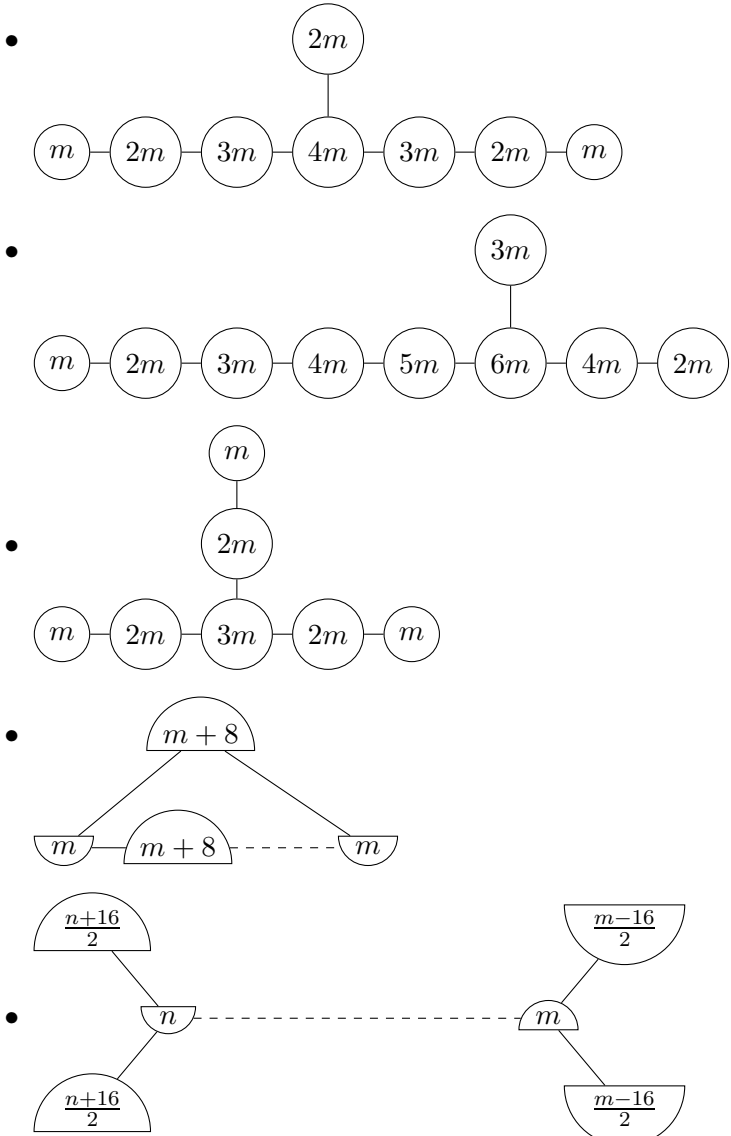
\begin{itemize}
\item\begin{tikzpicture}
\node[SU] (A) at  (4,1.5)  {$m$};
\node[SU] (B) at  (2,0)  {$m$};
\node[SU] (C) at  (3,0)  {$m$};
\node[SU] (D) at  (5,0)  {$m$};
\node[SU] (E) at  (6,0)  {$m$};
\draw (A)--(B)--(C);
\draw (A)--(E);
\draw[dashed] (3.4,0)--(4.6,0);
\draw (D)--(E);
\end{tikzpicture}
\item\begin{tikzpicture}
\node[SU] (Z) at  (1,-1.2)  {$m$};
\node[SU] (A) at  (1,1.2)  {$m$};
\node[SU] (B) at  (2,0)  {$2m$};
\node[SU] (C) at  (3.2,0)  {$2m$};
\node[SU] (D) at  (5.2,0)  {$2m$};
\node[SU] (E) at  (6.2,1.2)  {$m$};
\node[SU] (Y) at  (6.2,-1.2)  {$m$};
\draw (A)--(B)--(C);
\draw (Z)--(B);
\draw[dashed] (3.7,0)--(4.7,0);
\draw (D)--(E);
\draw (D)--(Y);
\end{tikzpicture}
\item\begin{tikzpicture}
\node[SU] (X) at  (-1.5,0)  {$m$};
\node[SU] (Y) at  (-0.4,0)  {$2m$};
\node[SU] (Z) at  (0.8,0)  {$3m$};
\node[SU] (A) at  (2,1.5)  {$2m$};
\node[SU] (B) at  (2,0)  {$4m$};
\node[SU] (C) at  (3.2,0)  {$3m$};
\node[SU] (D) at  (4.4,0)  {$2m$};
\node[SU] (E) at  (5.5,0)  {$m$};
\draw (A)--(B)--(C)--(D)--(E);
\draw (X)--(Y)--(Z)--(B);
\end{tikzpicture}
\item\begin{tikzpicture}
\node[SU] (X) at  (-1.5,0)  {$m$};
\node[SU] (Y) at  (-0.4,0)  {$2m$};
\node[SU] (Z) at  (0.8,0)  {$3m$};
\node[SU] (A) at  (4.4,1.5)  {$3m$};
\node[SU] (B) at  (2,0)  {$4m$};
\node[SU] (C) at  (3.2,0)  {$5m$};
\node[SU] (D) at  (4.4,0)  {$6m$};
\node[SU] (E) at  (5.6,0)  {$4m$};
\node[SU] (F) at  (6.8,0)  {$2m$};
\draw (A)--(D)--(E)--(F);
\draw (X)--(Y)--(Z)--(B)--(C)--(D);
\end{tikzpicture}
\item\begin{tikzpicture}
\node[SU] (X) at  (-1.5,0)  {$m$};
\node[SU] (Y) at  (-0.4,0)  {$2m$};
\node[SU] (Z) at  (0.8,0)  {$3m$};
\node[SU] (A) at  (0.8,1.2)  {$2m$};
\node[SU] (B) at  (0.8,2.4)  {$m$};
\node[SU] (D) at  (2,0)  {$2m$};
\node[SU] (E) at  (3.1,0)  {$m$};
\draw (B)--(A)--(Z)--(D)--(E);
\draw (X)--(Y)--(Z);
\end{tikzpicture}
\item\begin{tikzpicture}
\node[SO] (A) at  (3.8,1.5)  {$m+8$};
\node[Sp] (B) at  (2,0)  {$m$};
\node[SO] (C) at  (3.5,0)  {$m+8$};
\node[Sp] (D) at  (6,0)  {$m$};
\draw (A)--(B)--(C);
\draw (A)--(D);
\draw[dashed] (C)--(D);
\end{tikzpicture}
\item\begin{tikzpicture}
\node[SO] (Y) at (-1,-1.2) {$\frac{n+16}{2}$};
\node[SO] (Z) at (-1,1.2) {$\frac{n+16}{2}$};
\node[Sp] (A) at (0,0) {$n$};
\node[SO] (B) at (5,0) {$m$};
\node[Sp] (C) at (6,-1.2) {$\frac{m-16}{2}$};
\node[Sp] (D) at (6,1.2) {$\frac{m-16}{2}$};
\draw (Y)--(A);
\draw (Z)--(A);
\draw[dashed] (A)--(B);
\draw (B)--(C);
\draw (B)--(D);
\end{tikzpicture}\\
where a current of $8$ flows in the straight chain from $\USp(n)$ to $\SO(m)$.
\item\begin{tikzpicture}
\node[SO] (A) at (-3.5,0) {$\frac{m+16}{3}$};
\node[Sp] (B) at (-1.5,0) {$\frac{2m-16}{3}$};
\node[SO] (C) at (0,0) {$m$};
\node[Sp] (D) at (1.5,0) {$\frac{2m-16}{3}$};
\node[SO] (E) at (3.5,0) {$\frac{m+16}{3}$};
\node[Sp] (P) at (0,1.5) {$\frac{2m-16}{3}$};
\node[SO] (Q) at (0,3) {$\frac{m+16}{3}$};
\draw (A)--(B)--(C)--(D)--(E);
\draw (Q)--(P)--(C);
\end{tikzpicture}
\item
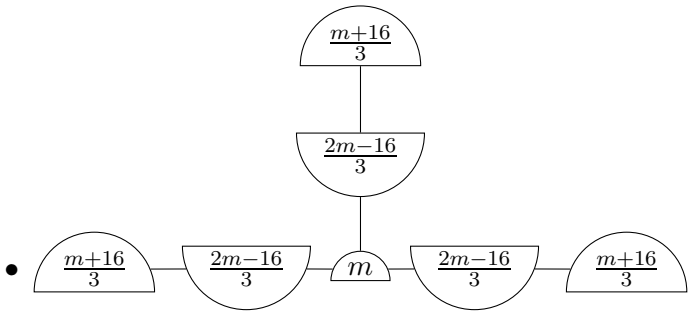
\begin{tikzpicture}
\node[Sp] (A) at (-3.5,0) {$\frac{n-16}{3}$};
\node[SO] (B) at (-1.5,0) {$\frac{2n+16}{3}$};
\node[Sp] (C) at (0,0) {$n$};
\node[SO] (D) at (1.5,0) {$\frac{2n+16}{3}$};
\node[Sp] (E) at (3.5,0) {$\frac{n-16}{3}$};
\node[SO] (P) at (0,1) {$\frac{2n+16}{3}$};
\node[Sp] (Q) at (0,2.5) {$\frac{n-16}{3}$};
\draw (A)--(B)--(C)--(D)--(E);
\draw (Q)--(P)--(C);
\end{tikzpicture}
\item\begin{tikzpicture}
\node[Sp] (Z) at (-3,0) {};
\node[SO] (A) at (-2,0) {};
\node[Sp] (B) at (-1,0) {};
\node[SO] (C) at (0,0) {$n$};
\node[Sp] (D) at (1,0) {};
\node[SO] (E) at (2,0) {};
\node[Sp] (F) at (3,0) {};
\node[Sp] (P) at (0,1) {};
\draw[->] (C)--node[right] {$i_u$}(P);
\draw[->] (C)--node[above] {$i_l$}(B);
\draw[->] (C)--node[below] {$i_r$}(D);
\draw (Z)--(A)--(B);
\draw (D)--(E)--(F);
\end{tikzpicture}\;\; Take three $\SO(n)$ branches of appropriate lengths with branch currents satisfying $i_u\le n/2$; $i_l,i_r\le n/4$ and $i_l+i_u+i_r\ge n-8$
\item\begin{tikzpicture}
\node[SO] (A) at (-2,0) {};
\node[Sp] (B) at (-1,0) {};
\node[SO] (C) at (0,0) {$n$};
\node[Sp] (D) at (1,0) {};
\node[SO] (E) at (2,0) {};
\node[Sp] (F) at (3,0) {};
\node[SO] (G) at (4,0) {};
\node[Sp] (H) at (5,0) {};
\node[Sp] (P) at (0,1) {};
\draw[->] (C)--node[right] {$i_u$}(P);
\draw[->] (C)--node[above] {$i_l$}(B);
\draw[->] (C)--node[below] {$i_r$}(D);
\draw (A)--(B);
\draw (D)--(E)--(F)--(G)--(H);
\end{tikzpicture}\; Take three $\SO(n)$ branches of appropriate lengths with branch currents satisfying $i_u\le n/2$; $i_l\le (n-8)/3$; $i_r\le n/6$ and $i_l+i_u+i_r\ge n-8$
\item
\begin{tikzpicture}
\node[SU] (A) at  (1,0) {};
\node[Sp] (B) at  (2,0) {$n$};
\node[SU] (C) at  (3,0) {};
\draw[->] (B)--node[above] {$i_l$}(A);
\draw[->] (B)--node[above] {$i_r$}(C);
\end{tikzpicture}\; $i_l, i_r\le n/2$; $i_l+i_r\ge n-8$   
\item
\begin{tikzpicture}
\node[SU] (A) at  (1,0) {};
\node[SU] (B) at  (2,0) {$n$};
\node[SU] (C) at  (3,0) {};
\node[Sp] (P) at  (2,1) {};
\draw[->] (B)--node[above] {$i_l$}(A);
\draw[->] (B)--node[below] {$i_r$}(C);
\draw[->] (B)--node[right] {$i_u$}(P);
\end{tikzpicture}\; $i_u\le8$; $i_l, i_r\le n/2$; $i_l+i_u+i_r\ge n$
\item
\begin{tikzpicture}
\node[SO] (Z) at  (0,0) {$m$};
\node[SU] (A) at  (1.5,0) {$m-8$};
\node[SU] (B) at  (6,0) {$n+8$};
\node[Sp] (C) at  (7.5,0) {$n$};
\draw (Z)--(A);
\draw[dashed] (A)--(B);
\draw (B)--(C);
\end{tikzpicture}\; A current of 8 flows from left to right 
\item
\begin{tikzpicture}
\node[Sp] (Z) at  (0,0) {$m$};
\node[SU] (A) at  (1,0) {};
\node[SU] (B) at  (4,0) {};
\node[Sp] (C) at  (5,0) {$n$};
\draw[->] (Z)--node[above] {$i$}(A);
\draw[dashed] (A)--(B);
\draw[->] (B)--node[above] {$j$}(C);
\end{tikzpicture}\; $i\ge-8$; $j\le8$; Take an $\SU(m-i)$ branch with branch current at least $i$ such that the current in last edge is $j$, and replace the last $\SU$ node by $\USp(n)$ with $8-j$ full hypers in $\fund$ of $\USp(n)$
\item
\begin{tikzpicture}
\node[Sp] (Z) at  (0,0) {};
\node[SU] (A) at  (1,0) {};
\node[SU] (B) at  (4,0) {};
\node[SU] (C) at  (5,0) {};
\node[SU] (D) at  (6,0) {$m$};
\node[SU] (P) at  (5,1) {};
\draw[->] (A)--node[above] {$j$}(Z);
\draw[dashed] (A)--(B);
\draw[->] (C)--node[right] {$i_u$}(P);
\draw[->] (C)--node[above] {$i_l$}(B);
\draw[->] (C)--node[below] {$i_r$}(D);
\end{tikzpicture}\; $j\le8$; Take three $\SU(m)$ branches (\emph{decorated} at the end by $\USp$ in the sense described in above point) with branch currents satisfying $i_u,i_r\le m/2$; $i_l+i_u+i_r\ge m$
\item
\begin{tikzpicture}
\node[SU] (A) at  (1,0) {};
\node[SU] (B) at  (2,0) {};
\node[SU] (C) at  (3,0) {};
\node[Sp] (D) at  (4,0) {$n$};
\node[SO] (E) at  (5,0) {};
\draw (A)--(B)--(C);
\draw[->] (D)--node[above] {$i_l$}(C);
\draw[->] (D)--node[above] {$i_r$}(E);
\end{tikzpicture}\; $i_r\le n/2$; $i_l\le n/4$; $2i_l+i_r\ge n-8$; and take an $\SU(n-i_l)$ branch of length 3 with branch current at least $i_l$
\item
\begin{tikzpicture}
\node[SU] (A) at  (1,0) {};
\node[SU] (B) at  (2,0) {};
\node[Sp] (C) at  (3,0) {$n$};
\node[SO] (D) at  (4,0) {};
\node[Sp] (E) at  (5,0) {};
\draw (A)--(B);
\draw[->] (C)--node[above] {$i_l$}(B);
\draw[->] (C)--node[above] {$i_r$}(D);
\draw (D)--(E);
\end{tikzpicture}\; Take a $\USp(n)$ branch of length three; $i_r\le (n+8)/3$; $i_l\le n/3$; $2i_l+i_r\ge n-8$; and take an $\SU(n-i_l)$ branch of length 2 with branch current at least $i_l$   
\item
\begin{tikzpicture}
\node[SO] (A) at  (1,0) {};
\node[Sp] (B) at  (2,0) {$n$};
\node[SO] (C) at  (3,0) {};
\node[SU] (P) at  (2,1) {};
\draw[->] (B)--node[above] {$i_l$}(A);
\draw[->] (B)--node[below] {$i_r$}(C);
\draw[->] (B)--node[right] {$i_u$}(P);
\end{tikzpicture}\; $i_l, i_u, i_r\le n/2$; $i_l+i_u+i_r\ge n$
\item
\begin{tikzpicture}
\node[SU] (Z) at  (0,0) {$m$};
\node[SO] (A) at  (1,0) {$2m$};
\node[Sp] (B) at  (4,0) {$2n$};
\node[SU] (C) at  (5,0) {$n$};
\draw (Z)--(A);
\draw[dashed] (A)--(B);
\draw (B)--(C);
\end{tikzpicture}\; A current of 8 flows from left to right in the alternating $\SO$-$\USp$ chain
\item
\begin{tikzpicture}
\node[SU] (Z) at  (0,0) {};
\node[Sp] (A) at  (1,0) {$m$};
\node[SO] (B) at  (2,0) {};
\node[SO] (C) at  (5,0) {};
\node[Sp] (D) at  (6,0) {$n$};
\node[SU] (E) at  (7,0) {};
\draw[->] (A)--node[above] {$i_l$}(Z);
\draw[->] (A)--node[above] {$i_r$}(B);
\draw[dashed] (B)--(C);
\draw[->] (D)--node[above] {$j_l$}(C);
\draw[->] (D)--node[above] {$j_r$}(E);
\end{tikzpicture}\; Take a $\USp(m)$ branch decorated at the end by $\SU$; $i_l\le m/2$; $j_r\le n/2$; $2i_l+i_r\ge m-8$; $2j_r+j_l\ge n-8$  
\item
\begin{tikzpicture}
\node[SU] (Z) at  (0,0) {$m$};
\node[SO] (A) at  (1,0) {$2m$};
\node[SO] (B) at  (4,0) {$2n$};
\node[Sp] (C) at  (5.5,0) {$n-8$};
\node[Sp] (P) at  (4,1) {$n-8$};
\draw (Z)--(A);
\draw[dashed] (A)--(B);
\draw (P)--(B)--(C);
\end{tikzpicture}\; A current of 8 flows in alternating $\SO$-$\USp$ chain from $\SO(2m)$ to $\SO(2n)$
\item
\begin{tikzpicture}
\node[SU] (Z) at  (0,0) {$m$};
\node[Sp] (A) at  (1,0) {$2m$};
\node[Sp] (B) at  (4,0) {$2n$};
\node[SO] (C) at  (5.5,0) {$n+8$};
\node[SO] (P) at  (4,1) {$n+8$};
\draw (Z)--(A);
\draw[dashed] (A)--(B);
\draw (P)--(B)--(C);
\end{tikzpicture}\; A current of 8 flows in alternating $\USp$-$\SO$ chain from $\USp(2n)$ to $\USp(2m)$
\item
\begin{tikzpicture}
\node[SU] (Z) at  (0,0) {};
\node[Sp] (A) at  (1,0) {$m$};
\node[SO] (B) at  (2,0) {};
\node[Sp] (C) at  (5,0) {};
\node[SO] (D) at  (6,0) {$n$};
\node[Sp] (E) at  (7,0) {};
\node[Sp] (P) at  (6,1) {};
\draw[->] (A)--node[above] {$j_l$}(Z);
\draw[->] (A)--node[above] {$j_r$}(B);
\draw[dashed] (B)--(C);
\draw[->] (D)--node[above] {$i_l$}(C);
\draw[->] (D)--node[below] {$i_r$}(E);
\draw[->] (D)--node[right] {$i_u$}(P);
\end{tikzpicture}\; Take a $\USp(n)$ branch with a decoration by $\SU$ at the end; $j_l\le m/2$; $i_u, i_r\le n/2$; $-j_r\ge i_l$; $i_l+i_u+i_r\ge n-8$; $2j_l+j_r\ge m-8$
\item
\begin{tikzpicture}
\node[Sp] (Y) at  (1,1) {};
\node[Sp] (Z) at  (0,0) {};
\node[SO] (A) at  (1,0) {$m$};
\node[Sp] (B) at  (2,0) {};
\node[Sp] (C) at  (5,0) {};
\node[SO] (D) at  (6,0) {$n$};
\node[Sp] (E) at  (7,0) {};
\node[Sp] (P) at  (6,1) {};
\draw[->] (A)--node[right] {$j_u$}(Y);
\draw[->] (A)--node[above] {$j_l$}(Z);
\draw[->] (A)--node[below] {$j_r$}(B);
\draw[dashed] (B)--(C);
\draw[->] (D)--node[above] {$i_l$}(C);
\draw[->] (D)--node[below] {$i_r$}(E);
\draw[->] (D)--node[right] {$i_u$}(P);
\end{tikzpicture}\; $j_l, j_u\le m/2$; $i_u, i_r\le n/2$; $-j_r\ge i_l$; $i_l+i_u+i_r\ge n-8$; $j_l+j_u+j_r\ge m-8$
\end{itemize}

\paragraph{Potential SCFTs:}
\begin{itemize}
\item Any $\SU$ branch
\item
\begin{tikzpicture}
\node[SU] (A) at  (1,0)  {};
\node[SU] (B) at  (4,0)  {};
\node[SU] (C) at  (5,0)  {$m$};
\node[SU] (D) at  (6,0)  {};
\node[SU] (P) at  (5,1)  {};
\draw[->] (C)--node[above] {$i_l$}(B);
\draw[->] (C)--node[below] {$i_r$}(D);
\draw[->] (C)--node[right] {$i_u$}(P);
\draw[dashed] (A)--(B);
\end{tikzpicture}\; Take an $\SU(m)$ branch of branch current $i_l$; $i_r, i_u\le m/2$; $i_l+i_u+i_r\ge m$
\item
\begin{tikzpicture}
\node[SU] (A) at  (1,0)  {};
\node[SU] (B) at  (2,0)  {};
\node[SU] (C) at  (3,0)  {$m$};
\node[SU] (D) at  (4,0)  {};
\node[SU] (E) at  (5,0)  {};
\node[SU] (P) at  (3,1)  {};
\draw[->] (C)--node[above] {$i_l$}(B);
\draw[->] (C)--node[below] {$i_r$}(D);
\draw[->] (C)--node[right] {$i_u$}(P);
\draw (A)--(B);
\draw (D)--(E);
\end{tikzpicture}\; Take three $\SU(m)$ branches of appropriate lengths; $i_r, i_l\le m/3$; $i_u\le m/2$; $i_l+i_u+i_r\ge m$
\item
\begin{tikzpicture}
\node[SU] (A) at  (1,0)  {};
\node[SU] (B) at  (2,0)  {};
\node[SU] (C) at  (3,0)  {$m$};
\node[SU] (D) at  (4,0)  {};
\node[SU] (E) at  (5,0)  {};
\node[SU] (F) at  (6,0)  {};
\node[SU] (P) at  (3,1)  {};
\draw[->] (C)--node[above] {$i_l$}(B);
\draw[->] (C)--node[below] {$i_r$}(D);
\draw[->] (C)--node[right] {$i_u$}(P);
\draw (A)--(B);
\draw (D)--(E)--(F);
\end{tikzpicture}\; Take three $\SU(m)$ branches of appropriate lengths; $i_r\le m/4$; $i_l\le m/3$; $i_u\le m/2$; $i_l+i_u+i_r\ge m$
\item
\begin{tikzpicture}
\node[SU] (A) at  (1,0)  {};
\node[SU] (B) at  (2,0)  {};
\node[SU] (C) at  (3,0)  {$m$};
\node[SU] (D) at  (4,0)  {};
\node[SU] (E) at  (5,0)  {};
\node[SU] (F) at  (6,0)  {};
\node[SU] (G) at  (7,0)  {};
\node[SU] (P) at  (3,1)  {};
\draw[->] (C)--node[above] {$i_l$}(B);
\draw[->] (C)--node[below] {$i_r$}(D);
\draw[->] (C)--node[right] {$i_u$}(P);
\draw (A)--(B);
\draw (D)--(E)--(F)--(G);
\end{tikzpicture}\; Take three $\SU(m)$ branches of appropriate lengths; $i_r\le m/5$; $i_l\le m/3$; $i_u\le m/2$; $i_l+i_u+i_r\ge m$
\item Any $\SO$ branch
\item Any $\USp$ branch
\item
\begin{tikzpicture}
\node[Sp] (B) at  (4,0)  {};
\node[SO] (C) at  (5,0)  {$m$};
\node[Sp] (D) at  (6,0)  {};
\node[Sp] (P) at  (5,1)  {};
\draw[->] (C)--node[above] {$i_l$}(B);
\draw[->] (C)--node[below] {$i_r$}(D);
\draw[->] (C)--node[right] {$i_u$}(P);
\draw[dashed] (1,0)--(B);
\end{tikzpicture}\; The dashed line denotes any $\SO(m)$ branch of branch current $i_l$; $i_r, i_u\le m/2$; $i_l+i_u+i_r\ge m-8$
\item
\begin{tikzpicture}
\node[SO] (B) at  (4,0)  {};
\node[Sp] (C) at  (5,0)  {$m$};
\node[SO] (D) at  (6,0)  {};
\node[SO] (P) at  (5,1)  {};
\draw[->] (C)--node[above] {$i_l$}(B);
\draw[->] (C)--node[below] {$i_r$}(D);
\draw[->] (C)--node[right] {$i_u$}(P);
\draw[dashed] (1,0)--(B);
\end{tikzpicture}\; The dashed line denotes a $\USp(m\ge16)$ branch of branch current $i_l$; $i_r, i_u\le m/2$; $i_l+i_u+i_r\ge m+8$
\item
\begin{tikzpicture}
\node[SO] (A) at  (1,0)  {};
\node[Sp] (B) at  (2,0)  {};
\node[SO] (C) at  (3,0)  {$m$};
\node[Sp] (D) at  (4,0)  {};
\node[SO] (E) at  (5,0)  {};
\node[Sp] (P) at  (3,1)  {};
\draw[->] (C)--node[above] {$i_l$}(B);
\draw[->] (C)--node[below] {$i_r$}(D);
\draw[->] (C)--node[right] {$i_u$}(P);
\draw (A)--(B);
\draw (D)--(E);
\end{tikzpicture}\; Take three $\SO(m)$ branches of appropriate lengths; $i_r, i_l\le (m-8)/3$; $i_u\le m/2$; $i_l+i_u+i_r\ge m-8$
\item
\begin{tikzpicture}
\node[Sp] (A) at  (1,0)  {};
\node[SO] (B) at  (2,0)  {};
\node[Sp] (C) at  (3,0)  {$m$};
\node[SO] (D) at  (4,0)  {};
\node[Sp] (E) at  (5,0)  {};
\node[SO] (P) at  (3,1)  {};
\draw[->] (C)--node[above] {$i_l$}(B);
\draw[->] (C)--node[below] {$i_r$}(D);
\draw[->] (C)--node[right] {$i_u$}(P);
\draw (A)--(B);
\draw (D)--(E);
\end{tikzpicture}\; Take three $\USp(m\ge32)$ branches of appropriate lengths; $i_r, i_l\le (m+8)/3$; $i_u\le m/2$; $i_l+i_u+i_r\ge m+8$
\item
\begin{tikzpicture}
\node[SO] (A) at  (1,0)  {};
\node[Sp] (B) at  (2,0)  {};
\node[SO] (C) at  (3,0)  {$m$};
\node[Sp] (D) at  (4,0)  {};
\node[SO] (E) at  (5,0)  {};
\node[Sp] (F) at  (6,0)  {};
\node[Sp] (P) at  (3,1)  {};
\draw[->] (C)--node[above] {$i_l$}(B);
\draw[->] (C)--node[below] {$i_r$}(D);
\draw[->] (C)--node[right] {$i_u$}(P);
\draw (A)--(B);
\draw (D)--(E)--(F);
\end{tikzpicture}\; Take three $\SO(m)$ branches of appropriate lengths; $i_r\le m/4$, $i_l\le (m-8)/3$; $i_u\le m/2$; $i_l+i_u+i_r\ge m-8$
\item
\begin{tikzpicture}
\node[Sp] (A) at  (1,0)  {};
\node[SO] (B) at  (2,0)  {};
\node[Sp] (C) at  (3,0)  {$m$};
\node[SO] (D) at  (4,0)  {};
\node[Sp] (E) at  (5,0)  {};
\node[SO] (F) at  (6,0)  {};
\node[SO] (P) at  (3,1)  {};
\draw[->] (C)--node[above] {$i_l$}(B);
\draw[->] (C)--node[below] {$i_r$}(D);
\draw[->] (C)--node[right] {$i_u$}(P);
\draw (A)--(B);
\draw (D)--(E)--(F);
\end{tikzpicture}\; Take three $\USp(m\ge64)$ branches of appropriate lengths; $i_r\le m/4$, $i_l\le (m+8)/3$; $i_u\le m/2$; $i_l+i_u+i_r\ge m+8$
\item
\begin{tikzpicture}
\node[SO] (A) at  (1,0)  {};
\node[Sp] (B) at  (2,0)  {};
\node[SO] (C) at  (3,0)  {$m$};
\node[Sp] (D) at  (4,0)  {};
\node[SO] (E) at  (5,0)  {};
\node[Sp] (F) at  (6,0)  {};
\node[SO] (G) at  (7,0)  {};
\node[Sp] (P) at  (3,1)  {};
\draw[->] (C)--node[above] {$i_l$}(B);
\draw[->] (C)--node[below] {$i_r$}(D);
\draw[->] (C)--node[right] {$i_u$}(P);
\draw (A)--(B);
\draw (D)--(E)--(F)--(G);
\end{tikzpicture}\; Take three $\SO(m)$ branches of appropriate lengths; $i_r\le (m-8)/5$, $i_l\le (m-8)/3$; $i_u\le m/2$; $i_l+i_u+i_r\ge m-8$
\item
\begin{tikzpicture}
\node[Sp] (A) at  (1,0)  {};
\node[SO] (B) at  (2,0)  {};
\node[Sp] (C) at  (3,0)  {$m$};
\node[SO] (D) at  (4,0)  {};
\node[Sp] (E) at  (5,0)  {};
\node[SO] (F) at  (6,0)  {};
\node[Sp] (G) at  (7,0)  {};
\node[SO] (P) at  (3,1)  {};
\draw[->] (C)--node[above] {$i_l$}(B);
\draw[->] (C)--node[below] {$i_r$}(D);
\draw[->] (C)--node[right] {$i_u$}(P);
\draw (A)--(B);
\draw (D)--(E)--(F)--(G);
\end{tikzpicture}\; Take three $\USp(m\ge112)$ branches of appropriate lengths; $i_r\le (m+8)/5$, $i_l\le (m+8)/3$; $i_u\le m/2$; $i_l+i_u+i_r\ge m+8$
\item
\begin{tikzpicture}
\node[SU] (A) at  (1,0) {};
\node[SU] (B) at  (4,0) {};
\node[SO] (C) at  (5,0) {$n$};
\draw[dashed] (A)--(B);
\draw[->] (C)--node[above] {$i$}(B);
\end{tikzpicture}\; $i\ge8$; Take an $\SU(n-i)$ branch with branch current of at least $i$
\item
\begin{tikzpicture}
\node[SU] (A) at  (1,0) {};
\node[SU] (B) at  (4,0) {};
\node[Sp] (C) at  (5,0) {$n$};
\draw[dashed] (A)--(B);
\draw[->] (C)--node[above] {$i$}(B);
\end{tikzpicture}\; $i\ge-8$; Take an $\SU(n-i)$ branch with branch current of at least $i$
\item
\begin{tikzpicture}
\node[SU] (B) at  (2,0) {};
\node[SU] (C) at  (3,0) {};
\node[Sp] (D) at  (4,0) {$n$};
\node[SO] (E) at  (5,0) {};
\draw (B)--(C);
\draw[->] (D)--node[above] {$i_l$}(C);
\draw[->] (D)--node[above] {$i_r$}(E);
\end{tikzpicture}\; $i_r\le n/2$; $i_l\le n/3$; $2i_l+i_r\ge n-8$; and take an $\SU(n-i_l)$ branch of length 2 with branch current at least $i_l$
\item
\begin{tikzpicture}
\node[SU] (B) at  (2,0) {};
\node[SU] (C) at  (3,0) {};
\node[SO] (D) at  (4,0) {$n$};
\node[Sp] (E) at  (5,0) {};
\draw (B)--(C);
\draw[->] (D)--node[above] {$i_l$}(C);
\draw[->] (D)--node[above] {$i_r$}(E);
\end{tikzpicture}\; $n\ge48$; $i_r\le n/2$; $i_l\le n/3$; $2i_l+i_r\ge n+8$; and take an $\SU(n-i_l)$ branch of length 2 with branch current at least $i_l$
\item
\begin{tikzpicture}
\node[SU] (Z) at  (0,0) {};
\node[Sp] (A) at  (1,0) {$m$};
\node[SO] (B) at  (2,0) {};
\draw[->] (A)--node[above] {$i_l$}(Z);
\draw[->] (A)--node[above] {$i_r$}(B);
\draw[dashed] (B)--(5,0);
\end{tikzpicture}\; Take a $\USp(m)$ branch with a branch current $i_r\ge-8$; $i_l\le m/2$; $2i_l+i_r\ge m-8$
\item
\begin{tikzpicture}
\node[SU] (Z) at  (0,0) {};
\node[SO] (A) at  (1,0) {$m$};
\node[Sp] (B) at  (2,0) {};
\draw[->] (A)--node[above] {$i_l$}(Z);
\draw[->] (A)--node[above] {$i_r$}(B);
\draw[dashed] (B)--(5,0);
\end{tikzpicture}\; Take an $\SO(m)$ branch with a branch current $i_r\ge8$; $i_l\le m/2$; $2i_l+i_r\ge m+8$
\end{itemize}

\subsection{Adding other matter}
Thanks to the restrictions imposed by the vanishing of global anomaly, we can replace \begin{tikzpicture}
\node[SU] (Z) at  (0,0) {2};
\end{tikzpicture} in our classification by $\SU(2)$ having $n_2=4$, \begin{tikzpicture}
\node[Sp] (Z) at  (0,0) {2};
\end{tikzpicture} by $\SU(2)$ having $n_2=10$ and \begin{tikzpicture}
\node[SU] (Z) at  (0,0) {3};
\end{tikzpicture} by $\SU(3)$ having $n_3=6$. However, in the case of $\SU(3)$ having $n_3=12$, we need to look at those theories in which there exists at least one $\SO$ or $\USp$ node that couples to other groups only through full hypers. Then, we need to replace one or more of such nodes by an $\SU(3)$ node having $n_3=12$ and redo the labelling analysis. There are only a few such theories and the labellings can be found exactly as we found the labellings for above mentioned theories.

A lot of other matter can be incorporated into our classification above in a similar fashion
\begin{itemize}
	\item An $\SU(n)$ node with a hyper charged under $\asym$ behaves exactly as a $\USp(n)$ node. We just need to consider theories where at least one $\USp(n)$ couples to other groups only through full hypers. We don't have to redo the labelling analysis for this case. One might worry that $n$ is always even for $\USp(n)$, but it doesn't matter as we have not used this to constrain our analysis so far. So, our previous results can be taken and extended for $\USp(n)$ with odd $n$.
	\item An $\SU(n)$ node with a hyper charged under $\sym$ behaves exactly as an $\SO(n)$ node. We just need to consider theories where at least one $\SO(n)$ couples to other groups only through full hypers. Fortunately, we don't have to redo the labelling analysis for this case either.
	\item An $\SU(6)$ node with a half hyper charged under $\asymT$ behaves like an $\SO$ or $\USp$ node. We just need to consider theories where at least one $\SO$ or $\USp$ couples to other groups only through full hypers. The labelling analysis has to be redone for this case.
\end{itemize}

\noindent This leaves us with $\spinor$ of $\SO(7\le n\le14)$. We do not consider $\SO(8)$ in the analysis below as it has an extra restriction that number of hypers in $\vect$, $\spinor$ and $\conj$ must be same and we are not keeping track of the number of $\vect$ in our analysis. 

We concern ourselves with only the structure of the unlabeled quivers. As far as the classification of structures is concerned, \begin{tikzpicture}
\node[SO] (A) at  (0,0) {13};
\node[Sq] (B) at  (1.5,0) {$\half\spinor$};
\draw (A)--(B);
\end{tikzpicture}, \begin{tikzpicture}
\node[SO] (A) at  (0,0) {11,12};
\node[Sq] (B) at  (1.5,0) {$\spinor$};
\draw (A)--(B);
\end{tikzpicture}, \begin{tikzpicture}
\node[SO] (A) at  (0,0) {9,10};
\node[Sq] (B) at  (1.5,0) {$2\spinor$};
\draw (A)--(B);
\end{tikzpicture}, and \begin{tikzpicture}
\node[SO] (A) at  (0,0) {7};
\node[Sq] (B) at  (1.5,0) {$4\spinor$};
\draw (A)--(B);
\end{tikzpicture} behave in the same way. Here \begin{tikzpicture}
\node[SO] (A) at  (0,0) {$m$};
\node[Sq] (B) at  (1.5,0) {$n\spinor$};
\draw (A)--(B);
\end{tikzpicture} denotes $n$ full hypers charged under $\spinor$ of $\SO(m)$. We denote all of these by a new vertex \begin{tikzpicture}
\node[S] (A) at  (0,0) {U};
\end{tikzpicture}.

\begin{tikzpicture}
\node[SO] (A) at  (0,0) {11,12};
\node[Sq] (B) at  (1.5,0) {$\half\spinor$};
\draw (A)--(B);
\end{tikzpicture}, \begin{tikzpicture}
\node[SO] (A) at  (0,0) {9,10};
\node[Sq] (B) at  (1.5,0) {$\spinor$};
\draw (A)--(B);
\end{tikzpicture}, \begin{tikzpicture}
\node[SO] (A) at  (0,0) {7};
\node[Sq] (B) at  (1.5,0) {$2\spinor$};
\draw (A)--(B);
\end{tikzpicture} also behave in the same way and we denote them by the vertex \begin{tikzpicture}
\node[S] (A) at  (0,0) {};
\end{tikzpicture}. Notice that adding just one hyper in $\spinor$ of $\SO(7)$ is not possible (unless it couples to other groups with hypers in $\spinor\otimes\fund$) because we cannot make $\alpha^{\SO(7)}=0$ by adding any number of hypers in fundamentals and binfundamentals.

We obtain some potential SCFTs:

\begin{itemize}
\item
\begin{tikzpicture}
\node[S] (Z) at  (0,0) {U};
\node[Sp] (A) at  (1,0) {};
\draw (Z)--(A);
\draw[dashed] (A)--(4,0);
\end{tikzpicture}\; where the dashed line denotes a chain of alternating $\USp$-$\SO$ nodes
\item
\begin{tikzpicture}
\node[S] (Z) at  (0,0) {};
\node[SU] (A) at  (1,0) {};
\draw (Z)--(A);
\draw[dashed] (A)--(4,0);
\end{tikzpicture}\; where the dashed line denotes a chain of $\SU$ nodes of length at most 2
\item
\begin{tikzpicture}
\node[S] (Z) at  (0,0) {};
\node[Sp] (A) at  (1,0) {};
\draw (Z)--(A);
\draw[dashed] (A)--(4,0);
\end{tikzpicture}
\item
\begin{tikzpicture}
\node[S] (Z) at  (0,0) {};
\node[Sp] (A) at  (1,0) {};
\node[SO] (P) at  (1,1) {};
\draw (Z)--(A)--(P);
\draw[dashed] (A)--(4,0);
\end{tikzpicture}\; where the total number of $\USp$ and $\SO$ nodes can at most be 6
\item
\begin{tikzpicture}
\node[S] (Z) at  (0,0) {};
\node[Sp] (A) at  (1,0) {};
\node[Sq] (B) at  (4,0) {};
\node[Sq] (C) at  (5,0) {};
\node[Sq] (P) at  (4,1) {};
\draw (Z)--(A);
\draw[dashed] (A)--(B);
\draw (P)--(B)--(C);
\end{tikzpicture}\; where a box means either a $\USp$ node or an $\SO$ node (still respecting the alternating condition)   
\item
\begin{tikzpicture}
\node[S] (Z) at  (0,0) {};
\node[Sp] (A) at  (1,0) {};
\node[SO] (B) at  (2,0) {};
\node[Sp] (C) at  (3,0) {};
\node[SO] (D) at  (4,0) {};
\node[Sp] (P) at  (2,1) {};
\draw (Z)--(A)--(B);
\draw (P)--(B)--(C)--(D);
\end{tikzpicture}
\item
\begin{tikzpicture}
\node[S] (Y) at  (-1,0) {};
\node[Sp] (Z) at  (0,0) {};
\node[SO] (A) at  (1,0) {};
\node[Sp] (B) at  (2,0) {};
\node[SO] (C) at  (3,0) {};
\node[Sp] (D) at  (4,0) {};
\node[SO] (P) at  (2,1) {};
\draw (Y)--(Z)--(A)--(B);
\draw (P)--(B)--(C)--(D);
\end{tikzpicture}
\item
\begin{tikzpicture}
\node[S] (Z) at  (0,0) {};
\node[Sp] (A) at  (1,0) {};
\node[Sq] (B) at  (4,0) {};
\node[SU] (C) at  (5,0) {};
\draw (Z)--(A);
\draw[dashed] (A)--(B);
\draw (B)--(C);
\end{tikzpicture}
\item
\begin{tikzpicture}
\node[S] (Z) at  (0,0) {};
\node[Sp] (A) at  (1,0) {};
\node[Sp] (B) at  (4,0) {};
\node[S] (C) at  (5,0) {};
\draw (Z)--(A);
\draw[dashed] (A)--(B);
\draw (B)--(C);
\end{tikzpicture}
\item
\begin{tikzpicture}
\node[S] (Z) at  (0,0) {};
\node[Sp] (A) at  (1,0) {};
\node[Sp] (B) at  (4,0) {};
\node[S] (C) at  (5,0) {U};
\draw (Z)--(A);
\draw[dashed] (A)--(B);
\draw (B)--(C);
\end{tikzpicture}
\item
\begin{tikzpicture}
\node[S] (Z) at  (0,0) {};
\node[Sp] (A) at  (1,0) {};
\node[SO] (C) at  (2,0) {};
\node[S] (P) at  (1,1) {};
\draw (Z)--(A);
\draw (P)--(A)--(C);
\end{tikzpicture}
\item
\begin{tikzpicture}
\node[S] (Z) at  (0,0) {};
\node[Sp] (A) at  (1,0) {};
\node[SO] (B) at  (2,0) {};
\node[Sp] (C) at  (3,0) {};
\node[SO] (D) at  (4,0) {};
\node[S] (P) at  (3,1) {};
\draw (Z)--(A)--(B)--(C);
\draw (P)--(C)--(D);
\end{tikzpicture}
\item
\begin{tikzpicture}
\node[Sp] (Y) at  (-1,0) {};
\node[S] (Z) at  (0,0) {};
\node[Sp] (A) at  (1,0) {};
\draw (Y)--(Z)--(A);
\draw[dashed] (A)--(4,0);
\end{tikzpicture}
\item
\begin{tikzpicture}
\node[SO] (X) at  (-2,0) {};
\node[Sp] (Y) at  (-1,0) {};
\node[S] (Z) at  (0,0) {};
\node[Sp] (A) at  (1,0) {};
\draw (X)--(Y)--(Z)--(A);
\draw[dashed] (A)--(4,0);
\end{tikzpicture}\; where the $\USp$-$\SO$ chain at the end must be composed of at most 4 nodes
\item
\begin{tikzpicture}
\node[Sp] (Y) at  (-1,0) {};
\node[S] (Z) at  (0,0) {};
\node[Sp] (A) at  (1,0) {};
\node[Sp] (B) at  (4,0) {};
\node[S] (C) at  (5,0) {};
\draw (Y)--(Z)--(A);
\draw[dashed] (A)--(B);
\draw (B)--(C);
\end{tikzpicture}
\item
\begin{tikzpicture}
\node[SO] (X) at  (-2,0) {};
\node[Sp] (Y) at  (-1,0) {};
\node[S] (Z) at  (0,0) {};
\node[Sp] (A) at  (1,0) {};
\node[S] (B) at  (2,0) {};
\draw (X)--(Y)--(Z)--(A)--(B);
\end{tikzpicture}
\end{itemize}

\noindent And we obtain some potential little string theories:

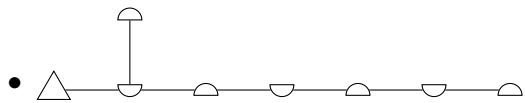
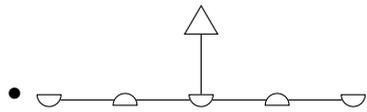
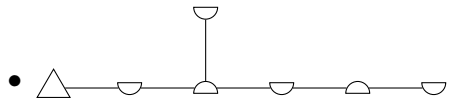
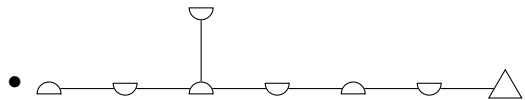
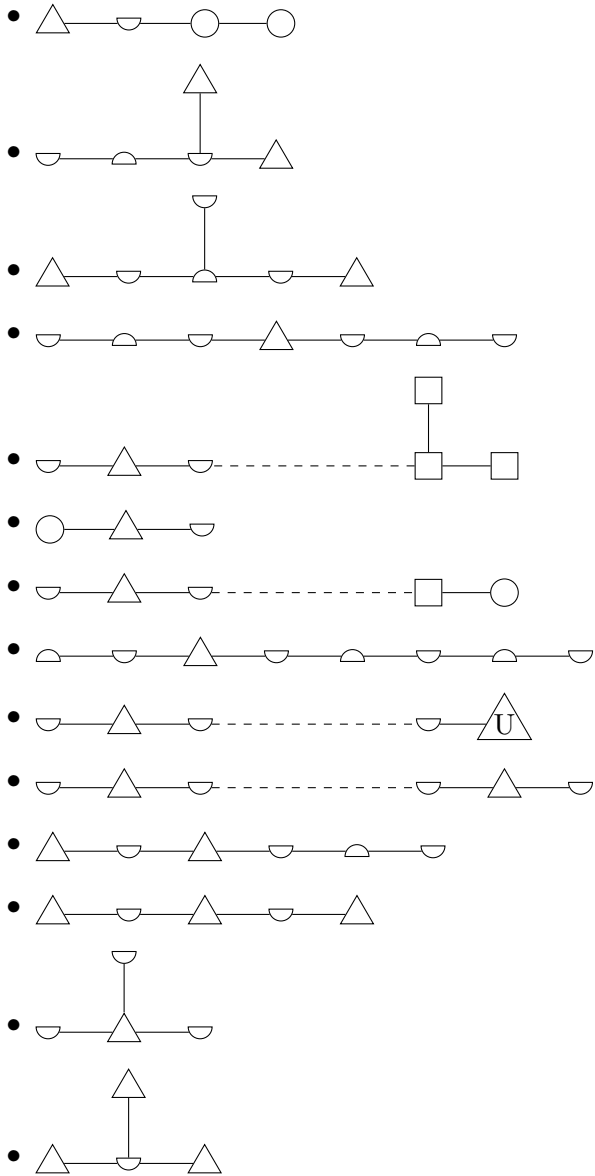
\begin{itemize}
\item
\begin{tikzpicture}
\node[S] (Z) at  (0,0) {U};
\node[Sp] (A) at  (1,0) {};
\node[Sp] (B) at  (4,0) {};
\node[S] (C) at  (5,0) {U};
\draw (Z)--(A);
\draw[dashed] (A)--(B);
\draw (B)--(C);
\end{tikzpicture}
\item
\begin{tikzpicture}
\node[S] (Z) at  (0,0) {U};
\node[Sp] (A) at  (1,0) {};
\node[Sq] (B) at  (4,0) {};
\node[Sq] (C) at  (5,0) {};
\node[Sq] (P) at  (4,1) {};
\draw (Z)--(A);
\draw[dashed] (A)--(B);
\draw (P)--(B)--(C);
\end{tikzpicture}
\item
\begin{tikzpicture}
\node[S] (Z) at  (0,0) {U};
\node[Sp] (A) at  (1,0) {};
\node[Sq] (B) at  (4,0) {};
\node[SU] (C) at  (5,0) {};
\draw (Z)--(A);
\draw[dashed] (A)--(B);
\draw (B)--(C);
\end{tikzpicture}
\item
\begin{tikzpicture}
\node[S] (Z) at  (0,0) {};
\node[SU] (A) at  (1,0) {};
\node[SU] (B) at  (2,0) {};
\node[SU] (C) at  (3,0) {};
\draw (Z)--(A)--(B)--(C);
\end{tikzpicture}
\item
\begin{tikzpicture}
\node[S] (Z) at  (0,0) {};
\node[Sp] (A) at  (1,0) {};
\node[SO] (B) at  (2,0) {};
\node[Sp] (C) at  (3,0) {};
\node[SO] (D) at  (4,0) {};
\node[Sp] (E) at  (5,0) {};
\node[SO] (F) at  (6,0) {};
\node[SO] (P) at  (1,1) {};
\draw (Z)--(A)--(B)--(C)--(D)--(E)--(F);
\draw (P)--(A);
\end{tikzpicture}
\item
\begin{tikzpicture}
\node[Sp] (A) at  (1,0) {};
\node[SO] (B) at  (2,0) {};
\node[Sp] (C) at  (3,0) {};
\node[SO] (D) at  (4,0) {};
\node[Sp] (E) at  (5,0) {};
\node[S] (P) at  (3,1) {};
\draw (A)--(B)--(C)--(D)--(E);
\draw (P)--(C);
\end{tikzpicture}
\item
\begin{tikzpicture}
\node[S] (A) at  (1,0) {};
\node[Sp] (B) at  (2,0) {};
\node[SO] (C) at  (3,0) {};
\node[Sp] (D) at  (4,0) {};
\node[SO] (E) at  (5,0) {};
\node[Sp] (F) at  (6,0) {};
\node[Sp] (P) at  (3,1) {};
\draw (A)--(B)--(C)--(D)--(E)--(F);
\draw (P)--(C);
\end{tikzpicture}
\item
\begin{tikzpicture}
\node[SO] (A) at  (1,0) {};
\node[Sp] (B) at  (2,0) {};
\node[SO] (C) at  (3,0) {};
\node[Sp] (D) at  (4,0) {};
\node[SO] (E) at  (5,0) {};
\node[Sp] (F) at  (6,0) {};
\node[S] (G) at  (7,0) {};
\node[Sp] (P) at  (3,1) {};
\draw (A)--(B)--(C)--(D)--(E)--(F)--(G);
\draw (P)--(C);
\end{tikzpicture}
\item
\begin{tikzpicture}
\node[S] (Z) at  (0,0) {};
\node[Sp] (A) at  (1,0) {};
\node[SU] (B) at  (2,0) {};
\node[SU] (C) at  (3,0) {};
\draw (Z)--(A)--(B)--(C);
\end{tikzpicture}
\item
\begin{tikzpicture}
\node[Sp] (A) at  (1,0) {};
\node[SO] (B) at  (2,0) {};
\node[Sp] (C) at  (3,0) {};
\node[S] (D) at  (4,0) {};
\node[S] (P) at  (3,1) {};
\draw (A)--(B)--(C)--(D);
\draw (P)--(C);
\end{tikzpicture}
\item
\begin{tikzpicture}
\node[S] (A) at  (1,0) {};
\node[Sp] (B) at  (2,0) {};
\node[SO] (C) at  (3,0) {};
\node[Sp] (D) at  (4,0) {};
\node[S] (E) at  (5,0) {};
\node[Sp] (P) at  (3,1) {};
\draw (A)--(B)--(C)--(D)--(E);
\draw (P)--(C);
\end{tikzpicture}
\item
\begin{tikzpicture}
\node[Sp] (Z) at  (0,0) {};
\node[SO] (A) at  (1,0) {};
\node[Sp] (B) at  (2,0) {};
\node[S] (C) at  (3,0) {};
\node[Sp] (D) at  (4,0) {};
\node[SO] (E) at  (5,0) {};
\node[Sp] (F) at  (6,0) {};
\draw (Z)--(A)--(B)--(C)--(D)--(E)--(F);
\end{tikzpicture}
\item
\begin{tikzpicture}
\node[Sp] (Z) at  (0,0) {};
\node[S] (A) at  (1,0) {};
\node[Sp] (B) at  (2,0) {};
\node[Sq] (C) at  (5,0) {};
\node[Sq] (D) at  (6,0) {};
\node[Sq] (P) at  (5,1) {};
\draw (Z)--(A)--(B);
\draw[dashed] (C)--(B);
\draw (P)--(C)--(D);
\end{tikzpicture}
\item
\begin{tikzpicture}
\node[SU] (Z) at  (0,0) {};
\node[S] (A) at  (1,0) {};
\node[Sp] (B) at  (2,0) {};
\draw (Z)--(A)--(B);
\end{tikzpicture}
\item
\begin{tikzpicture}
\node[Sp] (Y) at  (-1,0) {};
\node[S] (Z) at  (0,0) {};
\node[Sp] (A) at  (1,0) {};
\node[Sq] (B) at  (4,0) {};
\node[SU] (C) at  (5,0) {};
\draw (Y)--(Z)--(A);
\draw[dashed] (A)--(B);
\draw (B)--(C);
\end{tikzpicture}
\item
\begin{tikzpicture}
\node[SO] (Z) at  (0,0) {};
\node[Sp] (A) at  (1,0) {};
\node[S] (B) at  (2,0) {};
\node[Sp] (C) at  (3,0) {};
\node[SO] (D) at  (4,0) {};
\node[Sp] (E) at  (5,0) {};
\node[SO] (F) at  (6,0) {};
\node[Sp] (G) at  (7,0) {};
\draw (Z)--(A)--(B)--(C)--(D)--(E)--(F)--(G);
\end{tikzpicture}
\item
\begin{tikzpicture}
\node[Sp] (Y) at  (-1,0) {};
\node[S] (Z) at  (0,0) {};
\node[Sp] (A) at  (1,0) {};
\node[Sp] (B) at  (4,0) {};
\node[S] (C) at  (5,0) {U};
\draw (Y)--(Z)--(A);
\draw[dashed] (A)--(B);
\draw (B)--(C);
\end{tikzpicture}
\item
\begin{tikzpicture}
\node[Sp] (Y) at  (-1,0) {};
\node[S] (Z) at  (0,0) {};
\node[Sp] (A) at  (1,0) {};
\node[Sp] (B) at  (4,0) {};
\node[S] (C) at  (5,0) {};
\node[Sp] (D) at  (6,0) {};
\draw (Y)--(Z)--(A);
\draw[dashed] (A)--(B);
\draw (B)--(C)--(D);
\end{tikzpicture}
\item
\begin{tikzpicture}
\node[S] (Z) at  (0,0) {};
\node[Sp] (A) at  (1,0) {};
\node[S] (B) at  (2,0) {};
\node[Sp] (C) at  (3,0) {};
\node[SO] (D) at  (4,0) {};
\node[Sp] (E) at  (5,0) {};
\draw (Z)--(A)--(B)--(C)--(D)--(E);
\end{tikzpicture}
\item
\begin{tikzpicture}
\node[S] (Z) at  (0,0) {};
\node[Sp] (A) at  (1,0) {};
\node[S] (B) at  (2,0) {};
\node[Sp] (C) at  (3,0) {};
\node[S] (D) at  (4,0) {};
\draw (Z)--(A)--(B)--(C)--(D);
\end{tikzpicture}
\item
\begin{tikzpicture}
\node[Sp] (A) at  (1,0) {};
\node[S] (B) at  (2,0) {};
\node[Sp] (C) at  (3,0) {};
\node[Sp] (P) at  (2,1) {};
\draw (A)--(B)--(C);
\draw (P)--(B);
\end{tikzpicture}
\item
\begin{tikzpicture}
\node[S] (A) at  (1,0) {};
\node[Sp] (B) at  (2,0) {};
\node[S] (C) at  (3,0) {};
\node[S] (P) at  (2,1) {};
\draw (A)--(B)--(C);
\draw (P)--(B);
\end{tikzpicture}
\end{itemize}

\noindent One can go on to include other entries from Table~\ref{2-gon} that we have not included yet by the same strategy. The structure can be classified by adding them one by one in all possible ways to the above mentioned theories and checking the resulting determinant. Once the structures have been classified, one can classify all labellings associated to a structure in terms of some inequalities.

\section{Summary and future directions} \label{5}
We classified a large class of 6d $\cN=(1,0)$ gauge theories satisfying the consistency conditions that gauge anomalies can be cancelled by Green-Schwarz mechanism, the global anomaly vanishes and the charges of instanton strings in the theory are properly quantized. These theories fall into two classes
\begin{itemize}
	\item The number of tensor multiplets required to cancel the anomaly is equal to the number of gauge groups. We argued that if such theories have a UV completion it must be a 6d $\cN=(1,0)$ SCFT.
	\item The number of tensor multiplets required to cancel the anomaly is one less than the number of gauge groups. We argued that if such theories have a UV completion it must be a 6d $\cN=(1,0)$ little string theory.
\end{itemize}

One can ask which of the potential SCFTs we found have already been given a UV completion in F-theory \cite{Heckman:2015bfa,Heckman:2013pva}. If there are some potential SCFTs which don't appear in F-theory, there are a few possibilities to consider. First, these gauge theories could be pathological by themselves. For instance, they could violate some other consistency condition for 6d $\cN=(1,0)$ gauge theories. Second, they could have a UV completion in the form of a (1,0) SCFT but these SCFTs cannot be constructed in F-theory. Third, they could be consistent effective field theories in 6d but they don't come from deformations of (1,0) SCFTs. It would be very interesting to figure out which of the possibilities is true for theories which don't appear in F-theory constructions. Such developments can teach us about general properties of 6d $(1,0)$ theories.

The same set of questions can be asked for potential little string theories mentioned above. In a sense, these questions are more challenging and interesting because of a lack of known features/properties of little string theories. It would be nice to find precise arguments for/against our conjecture that 6d (1,0) gauge theories with one tensor multiplet less than the number of gauge groups must be little string theories if their UV completion exists.

It would also be interesting to understand the compactifications of these theories to lower dimensions and see what they can teach us about field theories with less than or equal to 8 supercharges in lower dimension and vice versa.

The close relation between the classifications of potential SCFTs and potential little string theories seems to suggest that there might exist a broader framework of 6d theories which treats 6d (1,0) SCFTs and little string theories almost on an equal footing. Understanding the relationship between the two might lead to new insights in the understanding of little string theories and may be also 6d SCFTs. 

\section*{Acknowledgements}
The author is grateful to Davide Gaiotto for suggesting the project. The author thanks Davide Gaiotto, Amihay Hanany and Yuji Tachikawa for useful discussions. The work of the author was supported by the Perimeter Institute for Theoretical Physics. Research at Perimeter Institute is supported by the Government of Canada through Industry Canada and by the Province of Ontario through the Ministry of Economic Development and Innovation.

\bibliographystyle{ytphys}
\let\bbb\bibitem\def\bibitem{\itemsep4pt\bbb}
\bibliography{ref}

\end{document}